\documentclass[sigconf]{acmart}
\AtBeginDocument{%
  }

\setcopyright{acmlicensed}
\copyrightyear{2018}
\acmYear{2018}
\acmDOI{}
\acmConference[Anonymous Conference]{Preprint}{}{}

\usepackage{url}
\usepackage{epsfig}
\usepackage[ruled,linesnumbered]{algorithm2e}
\usepackage{algpseudocode}
\usepackage{amsmath,amsthm}

\usepackage{amssymb}
\usepackage{xspace}
\usepackage{enumitem}
\usepackage{bbm}
\usepackage{cases}
\usepackage{color} 
\usepackage{newfloat}
\usepackage{listings}
\usepackage{multirow}
\usepackage{adjustbox}
\usepackage{booktabs,tabularx,booktabs}
\usepackage[ruled,linesnumbered]{algorithm2e}
\usepackage{algpseudocode}
\newcommand{\secfc}{SecPC\xspace}
\newcommand{\scheme}{SecFPP\xspace}
\newtheorem{theorem}{\bf{Theorem}} 

\theoremstyle{remark}

\DeclareMathOperator*{\argmin}{arg\,min}
\usepackage{rotating}
\usepackage{caption}
\usepackage{subcaption}
\usepackage{pifont}
\newcommand{\cmark}{\ding{51}}%
\newcommand{\xmark}{\ding{55}}%

\begin{document}

\title{Privacy-preserving Prompt Personalization in Federated Learning for Multimodal Large Language Models}


\author{Sizai Hou}
\email{shouac@connect.ust.hk}
\affiliation{%
  \institution{Hong Kong University of Science and Technology}
  \city{Hong Kong}
  \country{China}
}

\author{Songze Li}
\email{songzeli@seu.edu.cn}
\affiliation{%
  \institution{Southeast University}
  \city{Nanjing}
  \country{China}
}

\author{Baturalp Buyukates}
\email{b.buyukates@bham.ac.uk}
\affiliation{%
  \institution{University of Birmingham}
  \city{Birmingham}
  \country{United Kingdom}
}

\begin{abstract}
Prompt learning is a crucial technique for adapting pre-trained multimodal language models (MLLMs) to user tasks. Federated prompt personalization (FPP) is further developed to address data heterogeneity and local overfitting, however, it exposes personalized prompts -- valuable intellectual assets -- to privacy risks like prompt stealing or membership inference attacks. Widely-adopted techniques like differential privacy add noise to prompts, whereas degrading personalization performance. We propose \scheme, a secure FPP protocol harmonizing generalization, personalization, and privacy guarantees. \scheme employs hierarchical prompt adaptation with domain-level and class-level components to handle multi-granular data imbalance. For privacy, it uses a novel secret-sharing-based adaptive clustering algorithm for domain-level adaptation while keeping class-level components private. While theoretically and empirically secure, \scheme achieves state-of-the-art accuracy under severe heterogeneity in data distribution. Extensive experiments show it significantly outperforms both non-private and privacy-preserving baselines, offering a superior privacy-performance trade-off.

\end{abstract}



\maketitle

\section{Introduction}

Multimodal large language models (MLLMs) have received significant attention in recent years due to their remarkable generalization capabilities and strong performance in downstream tasks across diverse applications. However, the effectiveness of pre-trained MLLMs in specific downstream tasks is often limited by task-specific data distributions and the need for localized optimization. Prompt tuning has emerged as one of the most effective techniques for improving the performance of pre-trained models~\cite{prompttuning1,prompttuning2,prompttuning3,prompttuning4}. Subsequent works~\cite{zhou2022learning, zhou2022conditional, chen2023unleashing, prompttuning5} extended prompt tuning to vision-language models (VLMs) by introducing learnable prompt mechanisms, allowing for more flexible and data-efficient adaptation to downstream tasks. As a lightweight adaptation strategy, prompt learning enables efficient and effective customization of pre-trained models for user-specific downstream tasks. To mitigate local data overfitting and preserve data privacy, researchers have integrated prompt learning into federated learning (FL) frameworks ~\cite{guo2023promptfl, zhao2023fedprompt, zhao2023fedprompt}. This integration allows users to collaboratively train prompt parameters while benefiting from the global data distribution. Later works progressively developed the concept of \emph{federated prompt personalization (FPP)}, which has emerged as a promising approach for adapting pre-trained models to individual user tasks through various personalization techniques~\cite{guo2023pfedprompt, li2024FedOTP, cui2024harmonizing, yang2023efficient, li2023visual}. 

As in conventional FL, merely keeping data on local devices does not guarantee privacy in FPP. Numerous studies have demonstrated that gradient information leakage in FL can lead to successful privacy attacks (e.g., \cite{geiping2020inverting, zhu2019deep, zhao2020idlg, li2022auditing, yang2023gradient, petrov2024dager, feng2024uncovering, du2024sok, zhang2024graphleak, vu2024analysis, das2025security}), and these threats are equally applicable to FPP. To make matters worse, learned prompts usually represent high-value assets, as they include both task-specific knowledge and potentially sensitive user information ~\cite{shen2024prompt, wu2024quantifying, edemacu2024privacy}. 
As in FL (e.g.,~\cite{dwork2006differential,  abadi2016deep, bonawitz2017practical, wei2020federated, shi2022just, du2023dp, xiao2025differential, demelius2025recent}), the most prevalent countermeasure against such attacks in FPP is the application of differential privacy (DP). Consequently, recent FPP studies have also adopted DP as a standard privacy-preserving solution~\cite{guo2023pfedprompt, tran2025privacy}. 
Unfortunately, unlike full machine learning models, the lightweight nature of prompts makes them especially vulnerable to even minor perturbations introduced by DP noise. Under stringent privacy constraints, this sensitivity often results in significant performance degradation. As empirically demonstrated by \cite{tran2025privacy}, DP noise can lead to performance drops of up to 25\% under a strict privacy budget, even under low-rank adaption techniques designed to mitigate this effect. 

To address the fundamental trade-off between prompt personalization performance and data privacy guarantees, we propose \scheme, a novel FPP protocol that achieves strong privacy protection without sacrificing personalization performance. Our protocol leverages a secret-sharing primitive, \emph{Lagrange coded computation (LCC)}~\cite{yu2019lagrange}, and introduces a privacy-preserving prompt clustering mechanism, \secfc. In \scheme, each user decomposes their prompt into two components: a local prompt retained on-device, and a global prompt collaboratively learned across users. This decomposition enables granularity-aware prompt adaptation that effectively handles two levels of data heterogeneity. Specifically, global prompts capture coarse-grained distribution shifts (e.g., cross-domain heterogeneity) that groups of users may have, while local prompts accommodate fine-grained distribution shifts (e.g., inter-label heterogeneity) unique to individual users.

Our contributions are summarized as follows:
\begin{itemize}
    \item We propose \scheme, a privacy-preserving protocol for FPP that achieves user-level adaptation through a hierarchical prompt structure built upon two key components: LCC and \secfc. This hierarchical prompt strategy utilizes a granularity-aware adaptation scheme that balances generalization and personalization across the user federation. The granularity-aware components of the prompt globally adapt to the domain-level heterogeneity while locally accommodating the class-level heterogeneity.
    \item We develop \secfc, a novel privacy-preserving adaptive clustering algorithm designed for prompt clustering, leveraged to enable effective domain-level adaptation in \scheme.
    \item We provide both theoretical and empirical security analyses of \scheme, demonstrating that the protocol offers strong privacy guarantees with negligible computational and communicational overheads.
    \item We conduct extensive experiments under diverse data heterogeneity scenarios.
    Results show that \scheme achieves state-of-the-art personalization performance, matching or even surpassing existing non-private methods, while significantly outperforming privacy-preserving baselines.
\end{itemize}

\begin{figure*}[th]
  \centering
  \includegraphics[width=\linewidth]{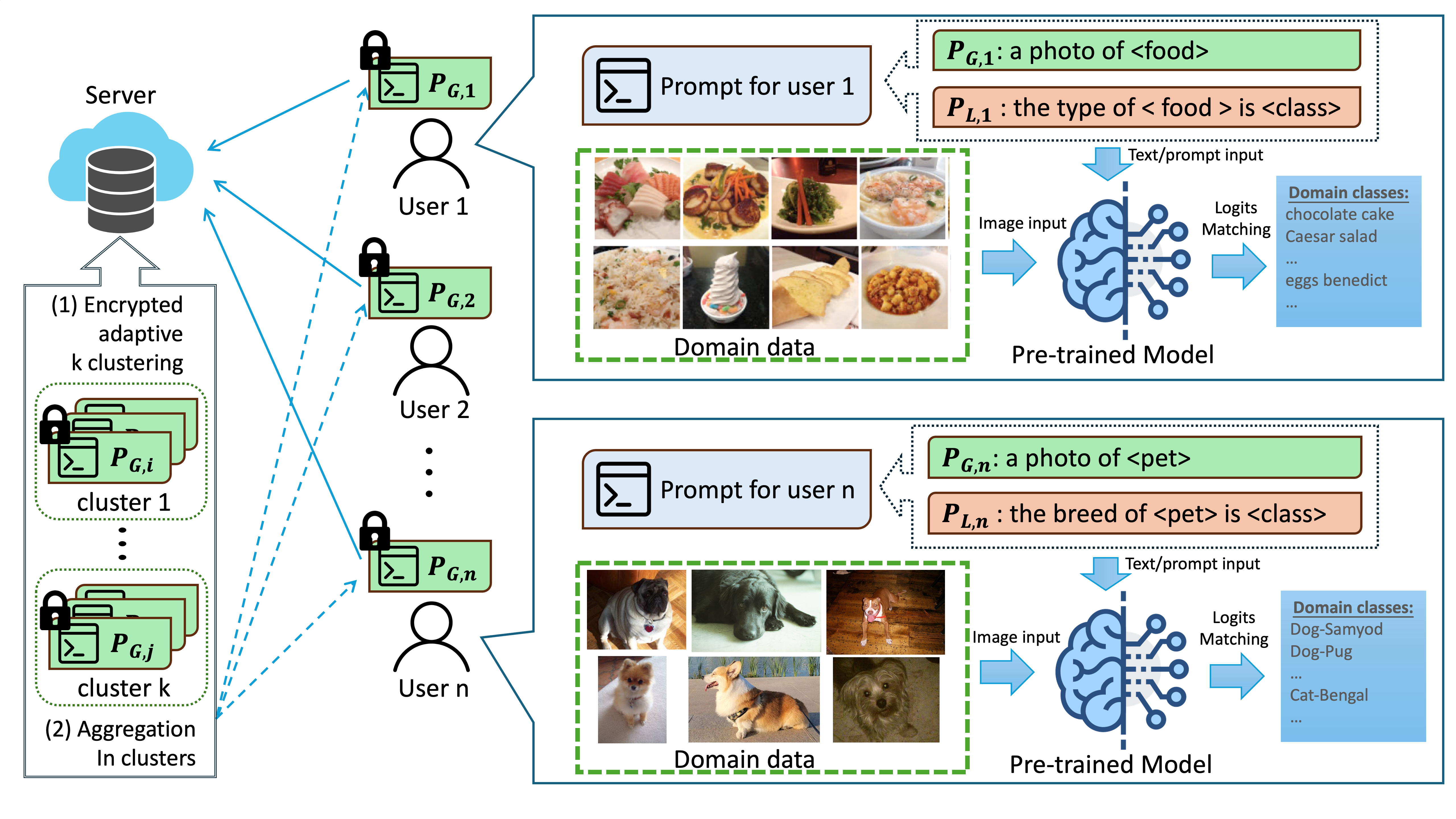}
  \vspace{-8mm}
  \caption{Workflow of \scheme. On the right, users decompose prompts into global and local components; the global prompt adapts to the dataset domain while the local prompt accommodates the local tasks. On the left, the global prompts are coded by Lagrange coded computation (LCC) and communicated among parties to enable adaptive clustering; then they are aggregated group-wise and distributed to users for the next training round.}
  \label{fig: workflow}
  \vspace{-3mm}
\end{figure*}

\begin{table}[t]
  \caption{Evolution of prompt learning. Hard prompts refer to manually engineered prompts, such as `\texttt{a photo of <class>}'. Soft prompts, introduced by CoOp~\cite{zhou2022learning, zhou2022conditional}, are learnable embeddings that were later integrated with the FL framework by PromptFL~\cite{guo2023promptfl} and Fed-Prompt~\cite{zhao2023fedprompt}.  Federated prompt personalization (FPP)~\cite{guo2023pfedprompt, cui2024harmonizing, li2024FedOTP} incorporates personalization techniques to adapt prompts and handle data heterogeneity. DP-FPL~\cite{tran2025privacy} further enhances the privacy of FPP by introducing differential privacy.}
  \label{tab:comparison}
  \begin{tabular}{lccc}
    \toprule
     & Single Dataset & Multi-Domain Dataset & Privacy  \\
    \midrule
    Hard Prompt & 68.2 & 61.9 & \xmark \\
    Soft Prompt & 91.4 & 84.2 & \xmark \\
    FPP & 91.6 & 85.5 & \xmark \\
    DP-FPL & 77.4 & 65.7 & \cmark \\
    \scheme (ours) & 91.6 & 91.2 & \cmark \\ 
  \bottomrule
\end{tabular}
\end{table}

\section{Related Works}
\subsection{Federated Prompt Personalization (FPP)}
First introduced for vision-language models by \cite{zhou2022conditional, zhou2022learning}, prompt learning is a lightweight and effective method to enhance the generalization capabilities of foundation models~\cite{prompttuning1,prompttuning2,prompttuning3,prompttuning4,prompttuning5,zhou2022conditional, zhou2022learning, chen2023unleashing}. Rather than relying on manually engineered prompts, prompt learning introduces parametric prompts, learnable continuous vectors trained on user data. Compared to computationally intensive full-model fine-tuning, prompt learning offers a more efficient alternative with notable gains in downstream performance as shown in Table \ref{tab:comparison}.
However, prompts trained solely on local data are prone to overfitting, especially in scenarios with limited or skewed user data distributions. To mitigate overfitting, users can employ federated collaborative prompt learning ~\cite{guo2023promptfl, zhao2023fedprompt}, jointly training prompt parameters while retaining data locally. The core challenge in this setting is to balance the foundation model’s generalization capacity with user-specific adaptation. Building on the concept of personalization in FL, recent studies have incorporated FL personalization techniques into prompt learning to better navigate this generalization and localization trade-off~\cite{zhao2023fedprompt, guo2023promptfl, guo2023pfedprompt, li2023visual, yang2023efficient, deng2024unlocking, li2024FedOTP, cui2024harmonizing}. 

Termed \emph{federated prompt personalization (FPP)}, this approach enables  efficient and effective local task adaptation for federated users while maintaining low computational and communication overhead. FedPrompt~\cite{zhao2023fedprompt} and PromptFL~\cite{guo2023promptfl} are the first to integrate FL into prompt learning through different FL paradigms. Subsequent works, such as pFedPrompt~\cite{guo2023pfedprompt} and pFedPG~\cite{yang2023efficient}, introduce more practical FPP methods to adapt frozen pre-trained models to local data heterogeneity. pFedPrompt uses a global prompt and an additional texture encoder for personalized attention to improve local task performance. pFedPG exploits a prompt generator at the server to provide personalized prompts for downstream users. Apart from textual prompts, \cite{li2023visual} proposes visual prompts that are attached to image inputs to represent local data distribution. \cite{deng2024unlocking} integrates FL model personalization with prompt selection techniques to resolve data heterogeneity effectively. FedOTP~\cite{li2024FedOTP} further improves the global prompt by user consensus knowledge extraction and uses the local prompt for capturing data features in severe data heterogeneity settings, such as label shifts and domain shifts. Addressing a similar challenge, FedPGP~\cite{cui2024harmonizing} adopts a low-rank prompt decomposition and additional contrastive loss to balance personalization and generalization.

\subsection{Privacy Preservation in FPP}
A personalized prompt is designed to guide a user's downstream task to fully exploit the capabilities of a pre-trained model. As user-specific customization becomes increasingly desirable, personalized prompts for commercially deployed LLMs are now widely recognized as valuable digital assets on various platforms such as OpenAI GPT Store~\cite{gptstore}, PromptBase~\cite{PromptBase}, SnackPrompt~\cite{snackprompt}. This growing ecosystem of prompt sharing and reuse introduces significant privacy risks. As high-value assets, customized user prompts for different model architectures have shown vulnerability to various attacks. Notably, prompt stealing attacks~\cite{shen2024prompt} and membership inference attacks~\cite{wu2024quantifying} have demonstrated the potential threats of unauthorized prompt usage, reproduction, or even leakage of proprietary or sensitive user data. In FL paradigms, a series of gradient-based privacy attacks have been demonstrated to successfully threaten user prompt information in FPP~\cite{geiping2020inverting, zhu2019deep, yang2023gradient, petrov2024dager, feng2024uncovering, du2024sok, zhang2024graphleak, das2025security}.

Although FPP has drawn considerable attention in recent research, its security risks remain rather under-explored, with only a few works addressing privacy concerns. PromptFL~\cite{guo2023promptfl} establishes a connection between differential privacy (DP) in FL and prompt learning. However, it does not effectively adapt DP to the prompt learning task and neglects the need for user-level personalization. A recent work, DP-FPL~\cite{tran2025privacy}, is the first to systematically propose a DP-based solution in FPP to seek a potential trade-off among personalization, generalization, and privacy. It implements global and local DP to protect prompts and leverages a low-rank adaption strategy to mitigate performance degradation induced by DP. Following prior work~\cite{cui2024harmonizing}, DP-FPL factorizes prompts into low-rank components to accommodate different data distributions while applying DP noise to low-rank components during training. Nevertheless, empirical results in Table~\ref{tab:comparison} present significant performance degradation (up to $25\%$ loss in certain cases) when a strict privacy budget is in place (e.g., $\epsilon \leq 0.01$). Additionally, the personalization design of DP-FPL falls short of adapting to domain-level data heterogeneity, as only the local prompt is factorized for personalization while the global prompt is learned universally.

\section{Problem Overview}
In this section, we present the problem formulation, followed by the threat model and overview of the proposed protocol, \scheme.

\subsection{Problem Formulation}
Let $F(\cdot)$ denote an upstream pre-trained model. In FPP, orchestrated by a central server, $n$ end users collaboratively train personalized prompts to adapt the pre-trained model to their datasets while fully utilizing the model's generalization capacity without overfitting. 
Denote user local datasets by $\left\{\mathcal{D}_1, \mathcal{D}_2, \ldots, \mathcal{D}_n\right\}$, where user $i$ has a distinct dataset $ \mathcal{D}_i$. Instead of using pre-defined prompts (e.g., “a photo of <label>”), users train their own prompts as learnable embeddings (or soft prompts), $\mathbf{P}_i \in \mathbb{R}^{d\times k}$, where $d$ is the dimension of the word embedding and $k$ is the number of tokens. We only keep the learnable parts in prompt and omit the hard prompt tokens with masked label positions for simplicity in formulations.

Consider data heterogeneity caused by different domains and an unbalanced split within a single domain. We denote personalized prompts by $\mathbf{P}_i$ and decompose it into global prompts and local prompts by $\mathbf{P}_i = {P}_{G,i}+P_{L,i}$. In the early training phase, the global prompt is shared across all users, i.e., ${P}_{G,i} = P_G$ for all $i$. During the later training stages, $P_{G,i}$ is trained group-wise according to the clustering assignment and aggregated inside the designated cluster (see Section \ref{workflow} for details). As a downstream task, each user $i$ performs classification as $\hat{y} = F(x, \mathbf{P}_i)$, for $\forall (x,y) \in \mathcal{D}_i$. Hence, the optimization objective for each user is given by:
\begin{equation}
     \mathcal{L}_{\mathcal{D}_i} = \sum_{(x,y)\in \mathcal{D}_i}\ell_{ce}(y , F(x, \mathbf{P}_i)),
    \label{eq: loss}
\end{equation}
where $\ell_{ce}$ stands for the cross entropy loss of the prediction $\hat{y}$ and $y$. The overall optimization problem for user prompts is:
\begin{equation}
    \argmin_{\{\mathbf{P}_i\mid i\in[n]\}}\sum_{i\in[n]}\frac{1}{\vert \mathcal{D}_i \vert}\mathcal{L}_{\mathcal{D}_i},
\end{equation}
where $[n]$ denotes the user index set $\{1,\ldots,n\}$ and $\vert \cdot \vert$ denotes the set cardinality.

\subsection{Threat Model}
The server and the users are \emph{honest-but-curious}, which is commonly used in evaluating the privacy of FL protocols~\cite{geiping2020inverting, bonawitz2017practical, so2022lightsecagg, buyukates2024lightverifl, tran2025privacy}. That is, all parties in the system follow the prescribed protocol faithfully, but the curious server and colluding users attempt to reveal private information including private prompts or membership, from both their local states and received messages during the protocol execution. We further assume that the server does not collude with the clients, as commonly considered in the secure federated learning literature~\cite{bonawitz2017practical, so2020byzantine, shao2022dres}. 

\subsection{Solution Overview}

Prior works in FPP mostly focus on personalization performance but overlook the security of the problem. Existing differentially private FPP works heavily suffer from model degradation and poor domain adaptation. 
To address the critical challenges in performance degradation and privacy preservation in FPP, we propose a secure federated prompt personalization protocol, \scheme, that integrates a privacy-preserving adaptive clustering algorithm and hierarchical prompt personalization scheme. In the ensuing sections, we first introduce the cryptographic building blocks including the coding primitive, LCC, and the privacy-preserving adaptive clustering algorithm, \secfc; then present the \scheme workflow in detail; and at last, provide the theoretical analysis for the security of the protocol.  The overview of \scheme is presented in Figure \ref{fig: workflow}.

\section{Proposed Secure FPP: \scheme}
This section presents the building blocks, workflow, and theoretical analysis of the proposed \scheme scheme.

\subsection{Cryptographic Building Blocks}\label{sec: building blocks}

\scheme involves an innovative privacy-preserving clustering algorithm, \secfc. Both clustering and aggregation are built upon the coding primitive LCC.

\noindent\textbf{LCC}. 
The key preliminary to achieve secure clustering is the secret-sharing scheme named Lagrange coded computing (LCC) ~\cite{yu2019lagrange}. LCC is a multi-secret sharing primitive, which generates shares of $\ell$ secrets $X_1,\ldots,X_{\ell}$ using Lagrange polynomial interpolation, where the shares are denoted by $[X_1,\ldots,X_{\ell}]_j$ for any shareholder $j$. LCC supports computations of $f(X_1),\ldots,f(X_{\ell})$, for any multi-variate polynomial $f(\cdot)$. The decryption of LCC is robust to missing and erroneous computation results with Reed-Solomon decoding. Compared with traditional Shamir secret sharing~\cite{shamir1979share} that encrypts one secret at a time, LCC reduces the share size and, thus, the load of computation on each party by $\ell$ times. Finally, we denote the usages of the sharing and reconstruction algorithms by $LCC.share(\cdot)$, $LCC.recon(\cdot)$, respectively. 

\noindent\textbf{\secfc}.
We propose a secure adaptive prompt clustering algorithm presented in Algorithm~\ref{alg: kmeans}, \secfc, which is developed upon the well-known $k$-means clustering algorithm and its adaptive variants~\cite{macqueen1967some, darken1990fast, bhatia2004adaptive, xia2020fast}. Since the prompt clustering is agnostic to the overall data distribution, the number of clusters is unknown initially. We build upon a simple yet effective adaptive $k$-means algorithm~\cite{bhatia2004adaptive} to perform clustering on user prompts federatively. During this clustering, only the relative distances to the cluster centers are communicated and revealed to the server, for which we present the security analysis in Section~\ref{sec: privacy thm}. Different from \cite{bhatia2004adaptive}, we do not update the centers whenever a data point is assigned. 
Instead, \secfc only performs cluster merging and center updating once at the end of each communication round. In this way, the clustering iterations correspond to the FL communication rounds (which can be considered as a one-shot clustering).

\noindent\textbf{Secure Aggregation}.
\scheme involves an aggregation step that group-wise aggregates global prompt components according to the cluster assignments $\mathcal{S}$. We employ the existing off-the-shelf secure aggregation protocols based on the same primitive of LCC, such as \cite{so2022lightsecagg, jahani2023swiftagg, buyukates2024lightverifl, hou2024priroagg}. 

\begin{algorithm}[tb]
\caption{\secfc : Secure Adaptive Clustering}
\label{alg: kmeans}
\KwData{Reduced local prompts $\overline{\mathbf{P}}_{i}$ of $n$ users, previous round cluster assignment $\mathcal{S}$.}
\KwOut{Updated $k$-cluster assignment $\mathcal{S} : \{ s_1,\ldots,s_k  \} $.}
Server broadcasts previous clustering $\mathcal{S}$ \;
\emph{Phase 1: Secure secret sharing}\;
\For{user $i \in [n]$ in parallel}{
    Slice its reduced local prompt $\overline{\mathbf{P}}_{i}$ by length $\ell$ and generate secret shares by LCC sharing scheme, $\left \{  \left [  \overline{\mathbf{P}}_{i}  \right ] _1, \ldots, \left [  \overline{\mathbf{P}}_{i}  \right ] _n \right \} \leftarrow LCC.Share(\overline{\mathbf{P}}_{i})$\;
    Share $\left [  \overline{\mathbf{P}}_{i}  \right ] _j$ with user $j\in[n]$\;
}
\emph{Phase 2: Homomorphic distance update}\;
\For{user $j \in [n]$ in parallel}{
    Update local coded center, $ [\mathbf{\mu_s}]_j \leftarrow \sum_{i\in s} \left [  \overline{\mathbf{P}}_{i}  \right ]_j, \forall s\in\mathcal{S}$ \;

    Compute coded distances for all users to each center $[d_{i,s}]_j \leftarrow \left \| [\mathbf{\mu_s}]_j - |s|\cdot \left [  \overline{\mathbf{P}}_{i}  \right ]_j \right \|_2^2, \quad\forall (i,s)\in[n]\times\mathcal{S}$\;
    Transmit all coded distances to the server\;
}
\emph{Phase 3: Cluster over revealed distances}\;
Server reconstructs each $d_{i,s}$ by Reed-Solomon decoding, $d_{i,s}\leftarrow LCC.recon(\{[d_{i,s}]_j \mid j\in[n]\}),\forall (i,s)\in[n]\times\mathcal{S}$ \;
Server recovers the real distances by $d_{i,s}\leftarrow \frac{d_{i,s}}{|s|^2},\forall (i,s)\in[n]\times\mathcal{S}$\;
Server performs one-shot adaptive clustering and updates cluster assignment $\mathcal{S}$.

\end{algorithm}

\subsection{\scheme Workflow}\label{workflow}

In the proposed \scheme scheme, we introduce an effective federated prompt personalization protocol that provides privacy guarantees while addressing multi-level data heterogeneity, presented in Figure \ref{fig: workflow}. In prior FPP works, various approaches are proposed to balance generality and local adaptation~\cite{li2024FedOTP,cui2024harmonizing,tran2025privacy}, but these methods all adopt a split structure of personalized prompt comprising a global prompt (shared across clients) and a local prompt (customized for individual users). 
In practice, data heterogeneity may come from diverse sources. It may arise from pathological class distributions within a single dataset but also may fundamentally come from distinct types of datasets.
Motivated by such split structure, we propose to use the splitting decomposition to achieve prompt adaptation for multi-level data heterogeneity: a global prompt component is dynamically adjusted to domain-level heterogeneity, and a local prompt component adapts local heterogeneity while remaining private on users to ensure privacy. To provide privacy preservation to domain-level adaption, we develop \secfc to achieve a simple yet effective approach to adaptively cluster the coarse-grained prompt adaptation of heterogeneity. 

At the beginning of training, the server initializes a universal prompt with global and local components for all users. In each communication round, each user computes the gradients for each component using loss $\mathcal{L}_{\mathcal{D}_i}$ measured on the local dataset $\mathcal{D}_i$. After finishing local training epochs, each user updates its local prompt. After updating the personalized prompt $\mathbf{P}_i$ by the local prompt, it performs dimensional reduction on the personalized prompt using a pre-defined method, such as the truncated SVD or PCA~\cite{wold1987principal}. The dimensional reduction preserves the most important components of the prompt while greatly reducing the transmitted data. The adaptive clustering, \secfc, is then performed federatively on the reduced $\overline{\mathbf{P}}_i$ and results in a cluster assignment, denoted by $\mathcal{S}$. According to the cluster assignment, the server invokes secure aggregation to group-wise aggregate the gradients for global components and back-propagates for the domain-level prompts. Finally, the global prompt component is updated into the personalized prompt for the next round or the downstream tasks. We summarize the protocol of \scheme in Algorithm \ref{alg: framework}. The detailed privacy-preserving clustering algorithm and aggregation schemes are explained in Sec.\ref{sec: building blocks}. 

\begin{algorithm}[tb]
\caption{\scheme: Secure Fed-Prompt Personalization}
\label{alg: framework}
\KwData{Local datasets $\left\{\mathcal{D}_1, \ldots, \mathcal{D}_n\right\}$ distributed on $n$ users}
\KwIn{Communication round $T$, learning rate $\eta$, decomposition rank $k$.}
\KwOut{Personalized prompts $\mathbf{P}_i$ for users}

Initialize cluster assignment $\mathcal{S} = \{[n]\}$\;
Initialize personalized prompt parameters $\mathbf{P}_i^{(0)} = P_{G,s}^{(0)} + P_{L,i}^{(0)}, s\in\mathcal{S}$\;
\For{iteration $t\leftarrow 0$ \KwTo $T-1$}{
\For{user $i \in [n]$ in parallel}{
Compute loss by Eq.\ref{eq: loss} and calculate gradients as $\nabla_{G,i} \mathcal{L}$ and $\nabla_{L,i} \mathcal{L}$ correspondingly\;
Update local prompt by $p_{L, i}^{(t+1)} \leftarrow p_{L, i}^{(t)}-\eta\nabla_{L, i} \mathcal{L}$\;
Use truncated SVD to reduce dimension for the personalized prompt $\overline{\mathbf{P}}_{i} \leftarrow PCA(\mathbf{P}_{i}^{(t+1)}, k)$ \;
}
Perform privacy-preserving adaptive $k$-means \secfc on reduced personalized prompts, $\mathcal{S}\leftarrow \secfc(\mathcal{S}, \{ \overline{\mathbf{P}}_{i} \mid i\in[n] \})$\;
Use secure aggregation for global prompt gradients according to cluster assignment, equivalently as, $\nabla_{G,s}\mathcal{L} \leftarrow \frac{1}{|s|}\sum_{i\in s}\nabla_{G,i} \mathcal{L}, \quad \forall s\in \mathcal{S}$\;
Update clustered global prompts by $P_{G,s}^{(t+1)} \leftarrow P_{G,s}^{(t)}-\eta\nabla_{G,s} \mathcal{L} ,\quad \forall s\in \mathcal{S}$\;
Users update personalized prompts by $\mathbf{P}_{i}^{(t+1)}\leftarrow P_{G,s\mid i\in s}^{(t+1)} + P_{L, i}^{(t+1)},\quad \forall i\in[n]$.
}
\end{algorithm}

\subsection{Theoretical Analysis}\label{sec: privacy thm}

As shown in Algorithms \ref{alg: kmeans} and \ref{alg: framework}, $P_{L,i}$ is always privately kept by users while $\overline{\mathbf{P}}_i, P_{G,s}$ join the secret-sharing based algorithms. LCC preserves perfect secrecy for the coded information within the security threshold of shareholders. Hence, we are interested in the theoretical analysis for the revealed information in the protocol, specifically, the reconstructed distances $d_{i,s}$, cluster assignment $\mathcal{S}$ and global prompts. Since cluster assignment is non-parametric and global prompts are considered public, we focus on the analysis of the reconstructed distances revealed to the honest-but-curious server. 

We assume that the prompt vector of each user has independent and identically distributed (i.i.d.) entries and satisfies a normal distribution of $\mathcal{N}(\mu_i, \sigma_i)$. We denote the vector space by $\mathcal{P}$. Without loss of generality, we analyze a single user prompt $\overline{\mathbf{P}}_i$ against a cluster of user prompts denoted by $\{\overline{\mathbf{P}}_1,\ldots,\overline{\mathbf{P}}_n\}$, such that $\overline{\mathbf{P}}_j \sim \mathcal{N}^d(\mu_j,\sigma_j), \forall j \in [n]$, where $i$ may or may not in $[n]$.  Prompts are independently sampled. We denote the average over the cluster by $\overline{\mathbf{P}}_{avg}$, which is the cluster center. To quantify the information leakage by the distance to the center, we analyze the mutual information between a prompt and the distance in the following theorem. 

\begin{theorem}\label{thm1}
    Given a cluster of prompts as normal random vectors by $\overline{\mathbf{P}}_j\sim \mathcal{N}^d(\mu_j,\sigma_j),j\in [n]$, the distance is the $\ell_2$-norm between a prompt $\overline{\mathbf{P}}_i$ and the cluster center $\overline{\mathbf{P}}_{avg}$, i.e., $D^2 = \left \| \overline{\mathbf{P}}_i -  \overline{\mathbf{P}}_{avg} \right \|^2_2$.
    The mutual information between $\overline{\mathbf{P}}_i$ and $D^2$ is given by:
    \begin{equation}\label{eq: thm1}\small
    \begin{split}
        MI\left(\overline{\mathbf{P}}_i;D^2\right) = & \log 2 \Gamma\left(\frac{d}{2}\right)+\left(1-\frac{d}{2}\right) \psi\left(\frac{d}{2}\right)+\frac{d}{2} \\  +\idotsint_{{\mathbf{P}} \in \mathcal{P}} & f_{\overline{\mathbf{P}}_i}\left(\overline{\mathbf{P}}_i={\mathbf{p}}\right) \cdot  \begin{cases}\left(\ln (2)+h_d\left(\frac{\tau}{2}\right)+c\right) \cdot d\mathbf{P}  & \text { if } d \in \mathbb{N}^{\text {odd }}, \\\left(\ln (2)+g_{d / 2}\left(\frac{\tau}{2}\right)+c\right) \cdot d\mathbf{P} & \text { if } d \in \mathbb{N}^{\text {even }} .\end{cases}
    \end{split}
    \end{equation}
    where $\Gamma, \psi$ represent gamma function and digamma function, respectively; $f_{\overline{\mathbf{P}}_i}$ is the probability density function of $\overline{\mathbf{P}}_i$; $c$ is $2\log\left(\frac{n-1}{n}\right)$; $h_n$ and $g_{n / 2}$ are two families of functions expanded in Appendix \ref{appdix: supp proof}. 
\end{theorem}
\begin{proof} (Part I)
     We are given reduced prompts $\{\overline{\mathbf{P}}_1,\ldots,\overline{\mathbf{P}}_n\}$ as d-dimensional continuous random vectors in a cluster, with sample mean $\overline{\mathbf{P}}_{avg} = \frac{1}{n}\sum_{j=1}^n \overline{\mathbf{P}}_j$. As all prompts satisfy $\overline{\mathbf{P}}_{j}\sim \mathcal{N}^d(\mu_j,\sigma_j)$, their average is also normally distributed, $\overline{\mathbf{P}}_{avg} \sim \mathcal{N}^d(\mu_{avg},\sigma_{avg})$, where $\mu_{avg},\sigma_{avg}^2$ are linear combinations of prompts' $\mu$ and $\sigma^2$. For a prompt $\overline{\mathbf{P}}_i$ inside or outside the given cluster, the distance between the prompt and the cluster center is $D^2 = \left \| \overline{\mathbf{P}}_i - \overline{\mathbf{P}}_{avg} \right \| ^2_2$. We discuss the inside case in this part of the proof and the outside case in Appendix \ref{appdix: supp proof} as Part II of proof. 
     
     Expand the mutual information in the form of differential entropy:
     \begin{equation}\label{eq: mi expansion to h}
         MI\left(\overline{\mathbf{P}}_i;D^2\right) = h\left(D^2\right) - h\left(D^2\mid\overline{\mathbf{P}}_i\right).
     \end{equation}
     For the entropy $h(D^2)$ we make the following observation: $D^2$ is simply a $\chi^2$-distribution with $d$ degrees of freedom, since it is the summation of the squares of all entries with centered normal distribution. Then:
     \begin{equation}\label{eq: entropy of distance}
         h(D^2) = \log 2 \Gamma\left(\frac{d}{2}\right)+\left(1-\frac{d}{2}\right) \psi\left(\frac{d}{2}\right)+\frac{d}{2},
     \end{equation}
     where $\Gamma$ and $\psi$ represent gamma function and digamma function, respectively.
     For the conditional entropy:
     \begin{align}
         h\left(D^2\mid\overline{\mathbf{P}}_i\right) =  h\left(\left\|\overline{\mathbf{P}}_i - \overline{\mathbf{P}}_{avg}\right\|^2_2 \mid \overline{\mathbf{P}}_i\right),
     \end{align}
     when $\overline{\mathbf{P}}_{avg}$ contains a term of $\overline{\mathbf{P}}_i$, by isolating it, we have:
     \begin{equation}\label{eq: conditional entropy expantion}
        \begin{split}
             h\left(D^2\mid\overline{\mathbf{P}}_i\right) & = h\left(\left \| \frac{n-1}{n}\overline{\mathbf{P}}_i - \frac{1}{n} \sum_{j=1,j\neq i}^{n} \overline{\mathbf{P}}_j \right \|^2_2\mid\overline{\mathbf{P}}_i\right) \\
             & = h\left(\left \| \overline{\mathbf{P}}_i - \frac{1}{n-1} \sum_{j=1,j\neq i}^{n} \overline{\mathbf{P}}_j \right \| ^2_2\mid\overline{\mathbf{P}}_i\right) + 2\log\left(\frac{n-1}{n}\right).
        \end{split}
     \end{equation}
    Notice $\overline{\mathbf{P}}_i$ and the normally distributed average, $\frac{1}{n-1} \sum_{j=1,j\neq i}^{n} \overline{\mathbf{P}}_j$, are independent. By taking the reduced prompt $\overline{\mathbf{P}}_i = {\mathbf{p}}$ as the conditional term, the first term of equation (\ref{eq: conditional entropy expantion}) is a non-central $\chi^2$-distribution with $d$ degrees of freedom and the non-centrality parameter $\tau$ is related to $\overline{\mathbf{P}}_i$ such that: 
     \begin{equation}\label{eq: tau}
             \tau \triangleq \sum_{k=1}^d \nu_{k}^2.
     \end{equation}
     We denote each entry of $\frac{1}{n-1} \sum_{j=1,j\neq i}^{n} \overline{\mathbf{P}}_j$ as $X_k$, such that the random variable $D^2$ is:
     \begin{equation}
         D^2 \triangleq \sum_{k=1}^d\left(X_k+\nu_k\right)^2.
     \end{equation}
     Then, $D^2$ is a non-central $\chi^2$-distribution with $d$ degrees of freedom and non-centrality parameter $\tau$.
     Apply the Theorem 1 in \cite{moser2020expected} using two families of function $g_m(\cdot), h_n(\cdot)$. We can reach the close form expression by:
     \begin{equation}
     \begin{split}
         h\left(\left \|\frac{1}{n-1} \sum_{j=1,j\neq i}^{n} \overline{\mathbf{P}}_j - {\mathbf{p}} \right \| ^2_2\mid\overline{\mathbf{P}}_i={\mathbf{p}}\right) 
         =
         f_{\mathbf{P}}\left(\overline{\mathbf{P}}_i={\mathbf{p}}\right)\mathbb{E}[-\log f(D^2)] \\
         = - f_{\mathbf{P}}\left(\overline{\mathbf{P}}_i={\mathbf{p}}\right)\cdot\begin{cases}\ln (2)+h_d\left(\frac{\tau}{2}\right)+c & \text { if } d \in \mathbb{N}^{\text {odd }}, \\ \ln (2)+g_{d / 2}\left(\frac{\tau}{2}\right)+c & \text { if } d \in \mathbb{N}^{\text {even }} .\end{cases} 
     \end{split}
     \end{equation}
     Hence, the overall conditional entropy is:
     \begin{equation}\label{eq: conditional entropy}
        \begin{split}
         h\left(\left \|\frac{1}{n-1} \sum_{j=1,j\neq i}^{n} \overline{\mathbf{P}}_j - \overline{\mathbf{P}}_i \right \| ^2_2\mid \overline{\mathbf{P}}_i \right)  = 
         \idotsint_{\mathbf{p} \in \mathcal{P}}  - f_{\overline{\mathbf{P}}_i}\left(\overline{\mathbf{P}}_i={\mathbf{p}}\right) \\ \cdot  \begin{cases}\left(\ln (2)+h_d\left(\frac{\tau}{2}\right)+c\right) \cdot d\mathbf{P}  & \text { if } d \in \mathbb{N}^{\text {odd }}, \\\left(\ln (2)+g_{d / 2}\left(\frac{\tau}{2}\right)+c\right) \cdot d\mathbf{P} & \text { if } d \in \mathbb{N}^{\text {even }} .\end{cases}
         \end{split}
     \end{equation}
     $f_{\overline{\mathbf{P}}_i}(\cdot)$ is the probability density function of $\overline{\mathbf{P}}_i$, i.e. a normal distribution; constant $c$ is $2\log\left(\frac{n-1}{n}\right)$. 
     Combining two terms of $h\left(D^2\right)$ and $ h\left(D^2\mid\overline{\mathbf{P}}_i\right)$, we have equation (\ref{eq: thm1}).  
         
\end{proof}

\noindent \textbf{Remark 1}: For rare edge cases, there is an infinitesimal possibility that one could reconstruct a user's entire prompt solely from distance measurements. However, when the prompts are randomly distributed, such reconstruction becomes statistically infeasible. Our focus is on typical scenarios and we aim to statistically answer the following question: \textit{To what extent can an honest-but-curious server infer information about a user’s prompt from reconstructed distances?} In Theorem \ref{thm1}, we provide an analytical formulation that characterizes the mutual information between a user's reduced prompt representation $\overline{\mathbf{P}}_i$ and the squared distance $D^2$, which quantifies the information leakage from the distance in a rigorous, statistical sense. Although deriving a tight upper bound on this mutual information is intractable, the expression serves as a theoretical foundation for analyzing privacy leakage. For instance, if $MI\left(\overline{\mathbf{P}}_i;D^2\right) << h\left(\overline{\mathbf{P}}_i\right)$, the distance provides negligible information about the prompt, i.e., observing $D^2$ reduces only an infinitesimal amount of uncertainty in the prompt. In such cases, the system satisfies information-theoretic privacy guarantees. To demonstrate that, we present empirical mutual information estimations in Section \ref{sec: privacy exp}.

\noindent \textbf{Remark 2}: Equation (\ref{eq: thm1}) has two main variables: $d$ and $n$. $MI(\overline{\mathbf{P}}_i;D^2)$ is dominated by the prompt dimension $d$ as the degrees of freedom. Though $n$ is a variable in the integral of constant, $\log\frac{n-1}{n}$ approaches zero when $n$ is large. The function family $h_n$ and $g_m$ exhibits logarithmic characteristics with strictly increasing monotonicity and is increasingly monotonic in $d$. In Section \ref{sec: privacy exp}, we numerically demonstrate these characteristics. 

\section{Experiments}
In this section, we evaluate the FPP performance of \scheme and compare it with private and non-private baselines under different levels of data heterogeneity. Moreover, we present empirical results for the mutual information estimations in the security analysis in Section \ref{sec: privacy thm}. Finally, we perform an evaluation of the computational overheads to demonstrate the cost of privacy in \scheme. 

\begin{table*}[tb]
  \caption{FPP performance with non-private and privacy-preserving protocols (accuracy in \%).} 
  \label{tab: main experiment}
  \begin{tabular}{cccccccccccccccc}
    \toprule
    &\multicolumn{6}{c}{Single Datasets} & \multicolumn{8}{c}{Multi-Domain Datasets} 
    \\\cmidrule(lr){2-7}\cmidrule(lr){8-15}
    \begin{tabular}{@{}c@{}} Personalization \\ Methods \end{tabular} & \begin{turn}{90} CIFAR-10 \end{turn} & \begin{turn}{90} Caltech-101 \end{turn} & \begin{turn}{90} Oxford-Pet \end{turn} & \begin{turn}{90} Oxford-Flowers \end{turn} & \begin{turn}{90} Food-101 \end{turn} & \begin{turn}{90} DTD \end{turn} & \begin{turn}{90} CIFAR-10+Caltech-101 \end{turn}   & \begin{turn}{90} CIFAR-100+Caltech-101 \end{turn}  & \begin{turn}{90} Caltech-101+Oxford-Pet \end{turn} & \begin{turn}{90} Caltech-101+Oxford-Flowers \end{turn} & \begin{turn}{90} Caltech-101+Food-101 \end{turn} & \begin{turn}{90} Caltech-101+DTD \end{turn} &  \begin{turn}{90} Oxford-Pet+Oxford-Flowers \end{turn} & \begin{turn}{90} Food-101+DTD \end{turn} \\
    \midrule
    PromptFL & 89.0 & 91.4 & 82.8 & 69.8 & 84.6 & 42.4 & 89.1 & 74.8 & 84.1 & 78.3 & 74.9 & 65.1 & 77.2 & 54.1\\
    FedOTP & \textbf{89.6} & \textbf{91.6} & \textbf{86.7} & 66.4  & \textbf{85.4} & \textbf{44.6} & \textbf{89.8} & 74.7 & 85.4 & \textbf{79.4} & 84.1 & \textbf{68.9}& 78.1 & \textbf{59.3}\\
    FedGPG & 88.9 & 91.1 & \textbf{86.8} & \textbf{70.1} & 84.7 & 42.6 & 88.6 & 74.0 & \textbf{91.4} & 76.3 & 75.5 & 67.8 & \textbf{81.2} & 57.8\\
    DP-FPL w.o. privacy  & 89.3 & 90.9 & 84.9 & \textbf{70.2} & 85.1 & 41.9 & 89.2 & 76.0 & 85.3 & 77.6 & 80.6 & 67.1 & 80.8 & 56.7\\
    
  \midrule
    DP-FPL w. loose privacy &  88.5 & 85.6 & 82.2 &   66.2  & 82.7 & 35.5 & 83.4 & 70.6 & 78.5 & 74.5 & 72.5 & 63.0 & 75.5 & 53.1\\
    DP-FPL w. default privacy & 86.1 & 77.4 & 77.6 & 61.2   & 72.3 & 32.6 & 82.9 & 62.2 & 73.1 & 53.1 &  66.3 & 45.4 & 64.0 & 49.1 \\
    DP-FPL w. strict privacy & 82.6 & 74.7 & 44.8 & 27.4   & 56.5 & 22.0 & 71.5  & 50.8 & 65.7 & 56.7 & 54.9 & 41.3 & 35.4  & 45.4 \\
  \midrule
    \scheme (ours) & \textbf{89.4} & \textbf{91.6} & \textbf{86.3} & \textbf{70.6} & \textbf{85.4} & \textbf{44.7} & \textbf{90.6} & \textbf{77.8} & \textbf{91.2} & \textbf{79.6} & \textbf{87.6} & \textbf{69.8} & \textbf{82.7} & \textbf{59.4}\\ 
  \bottomrule
\end{tabular}
\vspace{-3mm}
\end{table*}

\subsection{Experimental Settings}
Following previous works~\cite{tran2025privacy, cui2024harmonizing, li2024FedOTP, guo2023pfedprompt}, we perform the prompt-based image classification tasks on the pre-trained CLIP model~\cite{radford2021learning} using ViT-B/16 as backbone~\cite{dosovitskiy2020image}. 

\noindent\textbf{Datasets}. We consider various datasets from different domains to evaluate the FPP tasks. We use general-domain datasets: CIFAR-10, CIFAR-100~\cite{krizhevsky2009learning} and Caltech-101~\cite{li_andreeto_ranzato_perona_2022}; along with specific-domain datasets: Oxford-Pet~\cite{parkhi2012cats}, Oxford-Flowers~\cite{nilsback2008automated}, Food-101~\cite{bossard14}, and a texture database DTD~\cite{cimpoi14describing}. 
To comprehensively simulate data heterogeneity, we consider multi-granular heterogeneity for data distribution.
For single dataset allocation, we split the dataset evenly across users by Dirichlet distribution. Then, we allocate two domains of datasets to each half of the users. Within each dataset, we also apply Dirichlet distribution as the non-i.i.d. partition. This dual-level heterogeneity is denoted as datasetA + datasetB in the following section, e.g., Caltech101+OxfordPets. 

\noindent\textbf{Baselines}. For non-private baselines, we consider PromptFL~\cite{guo2023promptfl}, FedOTP~\cite{li2024FedOTP}, and FedGPG~\cite{cui2024harmonizing}. PromptFL is a federated version of CoOp~\cite{zhou2022conditional}. FedOTP and FedGPG are the existing state-of-the-art FPP protocols. Regarding privacy-preserving schemes, DP-enabled PromptFL is not capable of training personalized prompts, resulting in poor performance. Hence, we consider the only existing privacy-preserving FPP scheme, DP-FPL~\cite{tran2025privacy}, for comparisons. 

\noindent\textbf{Implementation details}. The default number of users, maximum number of epochs, number of local epochs, and learning rate are $20,100,10$, and $0.001$, respectively. For protocols involving low-rank decomposition, we set the default rank to $8$. In \scheme, we set the first-k principal components to $8$ as well. For DP-FPL, we choose DP parameter $\epsilon$ from $\{0.0, 0.4, 0.1, 0.01\}$ as no privacy, loose privacy, default privacy, and strict privacy constraints, respectively. We set the Dirichlet distribution parameter to $\beta=0.3$. 

\vspace{-3mm}
\subsection{FPP Performance Comparisons}

The overall performance is presented in Table \ref{tab: main experiment}. In general domain datasets, such as CIFAR-10, Caltech-101, all baselines have marginal differences. In contrast, for single specific-domain datasets, FPP-based methods consistently achieve higher accuracy compared to non-personalized PromptFL. \scheme also achieves comparably strong accuracy results in all single-domain datasets. 
In multi-domain datasets, the three non-private FPP approaches demonstrate distinct advantages on different dataset combinations while \scheme presents consistent advantages in accuracy. When the combined datasets are general (such as CIFAR-10+Caltech-101), all methods demonstrate comparable accuracy levels, although our solution exhibits a marginal but consistent performance advantage. Notably, in scenarios with high domain discrepancy, \scheme's advantage becomes particularly pronounced. While FedOTP demonstrates competitive robustness under severe data heterogeneity compared to other FPP approaches, \scheme consistently surpasses all existing FPP solutions in these challenging scenarios.

On the other hand, for the private FPP baselines, while DP-FPL can achieve comparable performance to non-private solutions without DP noise, its accuracy significantly degrades when DP is applied. Furthermore, under tighter privacy constraints, DP-FPL’s performance deteriorates proportionally. In sharp contrast, our \scheme provides rigorous privacy guarantees without compromising any model performance, while demonstrating superior robustness in personalization for both single-domain and multi-domain heterogeneous data scenarios.

\vspace{-3mm}
\subsection{Empirical Studies for Security}\label{sec: privacy exp}

\begin{figure}[tb]
\centering
    \centering
    \begin{subfigure}[b]{0.23\textwidth}
        \centering
        \includegraphics[width=\textwidth]{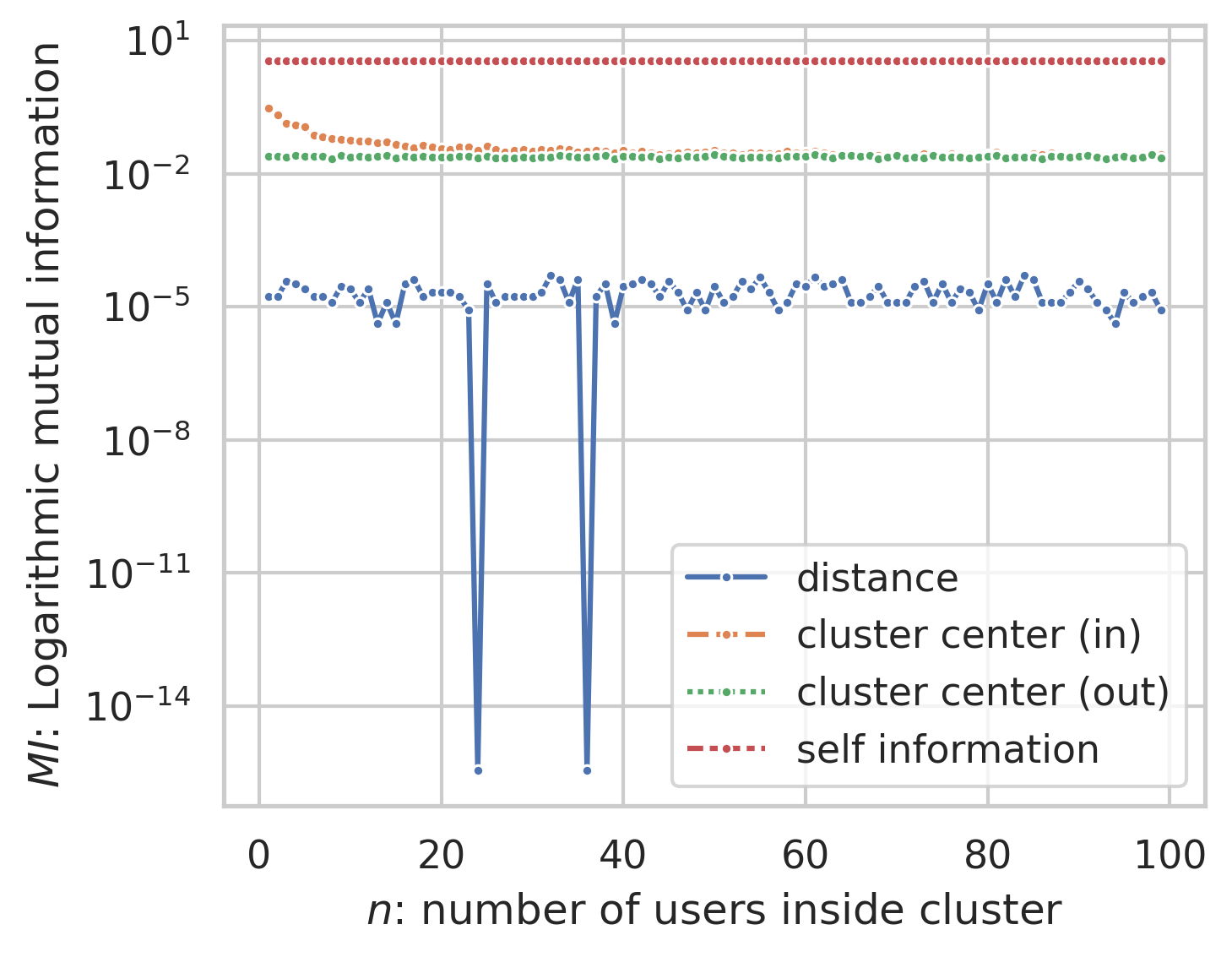}
        \caption{MI vs. user number}
        \label{fig: mi vs user num}
    \end{subfigure}
    \hfill
    \begin{subfigure}[b]{0.23\textwidth}
        \centering
        \includegraphics[width=\textwidth]{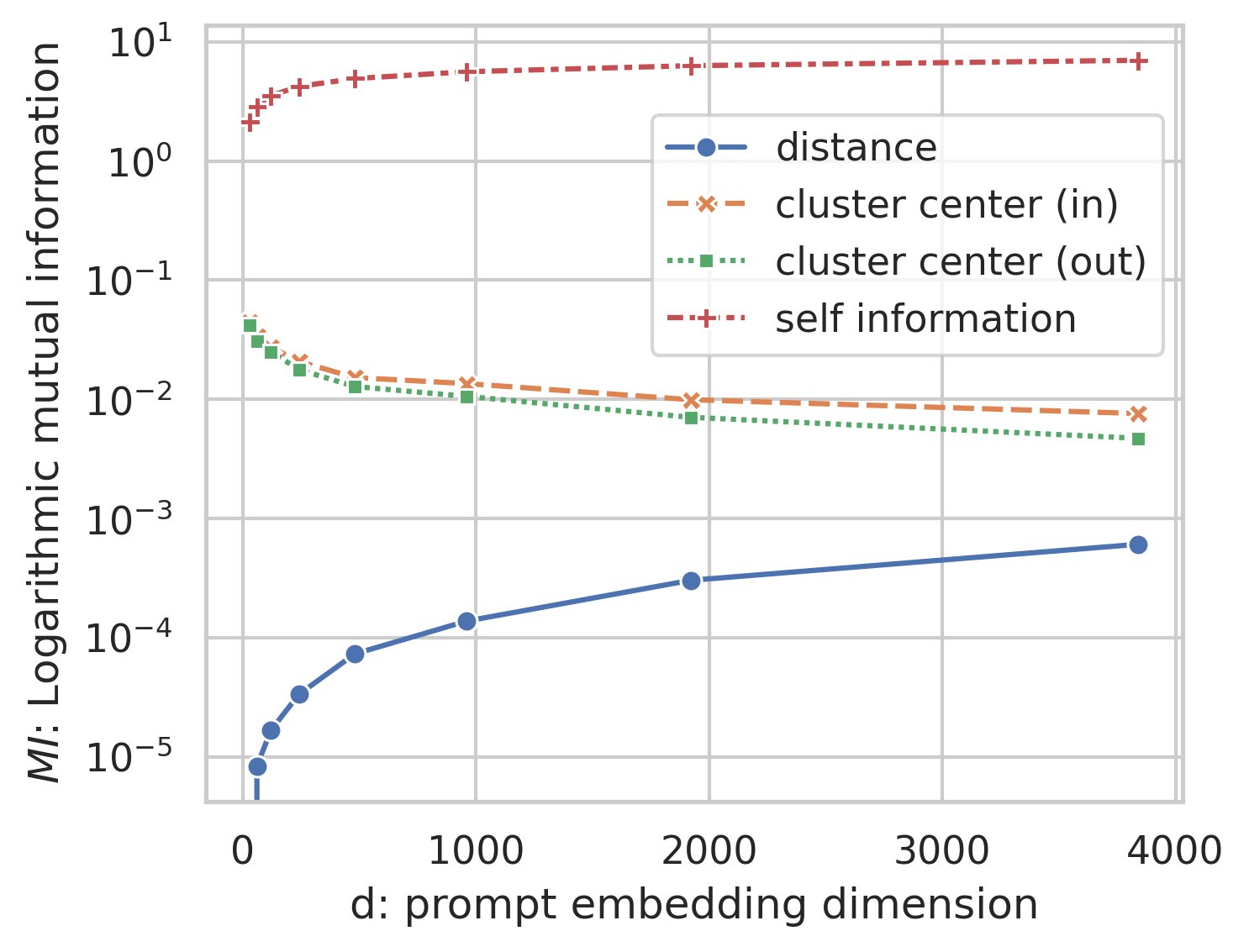}
        \caption{MI vs. prompt dimension}
        \label{fig: mi vs dim}
    \end{subfigure}
    \vspace{-3mm}
    \caption{The mutual information comparisons: distance stands for $MI( \overline{\mathbf{P}}_i; D^2 )$; cluster center is $MI( \overline{\mathbf{P}}_i; \overline{\mathbf{P}}_{avg} )$, which has two cases of the prompt being inside/outside the cluster; self-information is $h(\overline{\mathbf{P}}_i)$. To present the scale, mutual information is plotted in log-scale.}
    \label{fig: MI}
    \vspace{-3mm}
\end{figure}

In this subsection, we present numerical analysis for mutual information (MI) following the analytical results in Theorem \ref{thm1}. To evaluate it, we exploit widely used KSG algorithm~\cite{kraskov2004KSG} for estimations. Referring to earlier studies~\cite{ross2014mutual, gao2018demystifying, wang2021privacy}, we apply the MI estimation on independently sampled user prompts.
Recall in Section \ref{sec: privacy thm}, the main focus is on `\textit{how much information about the prompt an honest-but-curious server can infer from the reconstructed distances?}'. We aim to demonstrate the distance's statistical insignificance with respect to a reduced prompt. Hence, we consider three variables for estimating MI as comparisons: entropy of a user prompt (self MI), $h(\overline{\mathbf{P}}_i)$; MI between a user prompt and cluster center, $MI( \overline{\mathbf{P}}_i; \overline{\mathbf{P}}_{avg} )$; MI between a user prompt and the distance, $MI( \overline{\mathbf{P}}_i; D^2 )$. For $MI( \overline{\mathbf{P}}_i; \overline{\mathbf{P}}_{avg} )$, we consider both cases such that the prompt can be inside or outside the given cluster.  As in \cite{gao2018demystifying}, numerical analysis shows that a sample size over $1000$ reduces mean squared residuals to $10^{-3}$, ensuring stable estimation. So, we use $1000$ as sample size and sample user prompts as i.i.d. random vectors. Specifically, the prompt has $15$ tokens and the dimension is $512$ by default. The dimension is then reduced to $8$ as part of the \scheme, and $(\mu_i,\sigma_i)$s are simplified to $(0,1)$. For the distance as a number, we directly use the deterministic scalar-to-vector mapping (replicating $D^2$ across dimensions) to form a vector for MI estimation, such that the entropy is preserved by the deterministic function. The evaluation results are presented in Figure \ref{fig: MI}.

As demonstrated by the blue trajectory in Figures \ref{fig: mi vs user num} and \ref{fig: mi vs dim}, the prompt dimension $d$ is the predominant factor and exhibits a positive correlation with MI, while the user number $n$ has negligible influence (where $n$ is formally $|\mathcal{S}|$). The empirical results correctly align with the analytical result in Equation (\ref{eq: thm1}).
Across both evaluations, the mutual information $MI( \overline{\mathbf{P}}_i; D^2 )$ between the distance and the prompt (blue) remains exponentially lower than the prompt's entropy (red). Specifically, we have $\log_{10}\frac{h(\overline{\mathbf{P}}_i)}{MI( \overline{\mathbf{P}}_i; D^2 )}>r$, with $r$ serving as a lower bound. In our experiments, we observe $r=4$ such that $MI(\overline{\mathbf{P}}_i;D^2) << h(\overline{\mathbf{P}}_i)$. This substantial gap provides a rigorous information-theoretic constraint on any adversarial reconstruction attempts. Thus, \textit{\scheme offers strong information-theoretic privacy guarantees for user prompts in practical settings}.

We also present mutual information $MI( \overline{\mathbf{P}}_i; \overline{\mathbf{P}}_{avg} )$ between a given prompt and the cluster center for further illustrations. When a prompt resides within a cluster and the cluster cardinality is small, the cluster center exhibits higher statistical dependence on the prompt, inducing larger $MI( \overline{\mathbf{P}}_i; \overline{\mathbf{P}}_{avg} )$. The dependency diminishes asymptotically by increasing cluster cardinality $|\mathcal{S}|$ and dimension $d$. Generally, the observation agrees with the expectation that the cluster center vector conveys more information than the scalar distance while diluting the original prompt entropy by averaging over more prompts. Thus, the fact that \scheme only reveals relative distances, rather than explicit cluster centers, offers inherently stronger security guarantees, as it significantly limits the server's ability to infer sensitive prompt information.

\vspace{-2mm}
\subsection{Complexity Evaluations}

Here, we focus on the additional computation and communication costs of achieving privacy guarantees with \scheme. All experiments are conducted on a machine using Intel(R) Xeon(R) Gold 5118 CPU @ 2.30GHz, with 12 cores of 48 threads. The computational capability of a user is constrained to one-quarter of the server by parallelization. We denote the LCC parameters by $\ell$ and the privacy threshold by $t=\alpha\cdot n$, where $\alpha$ is a constant allowing privacy against $\left \lfloor \alpha\cdot n \right \rfloor $ colluding users. The complexity is summarized in Table \ref{tab: complexity}.

\begin{figure}[tbh]
\centering
    \centering
    \begin{subfigure}[b]{0.23\textwidth}
        \centering
        \includegraphics[width=\textwidth]{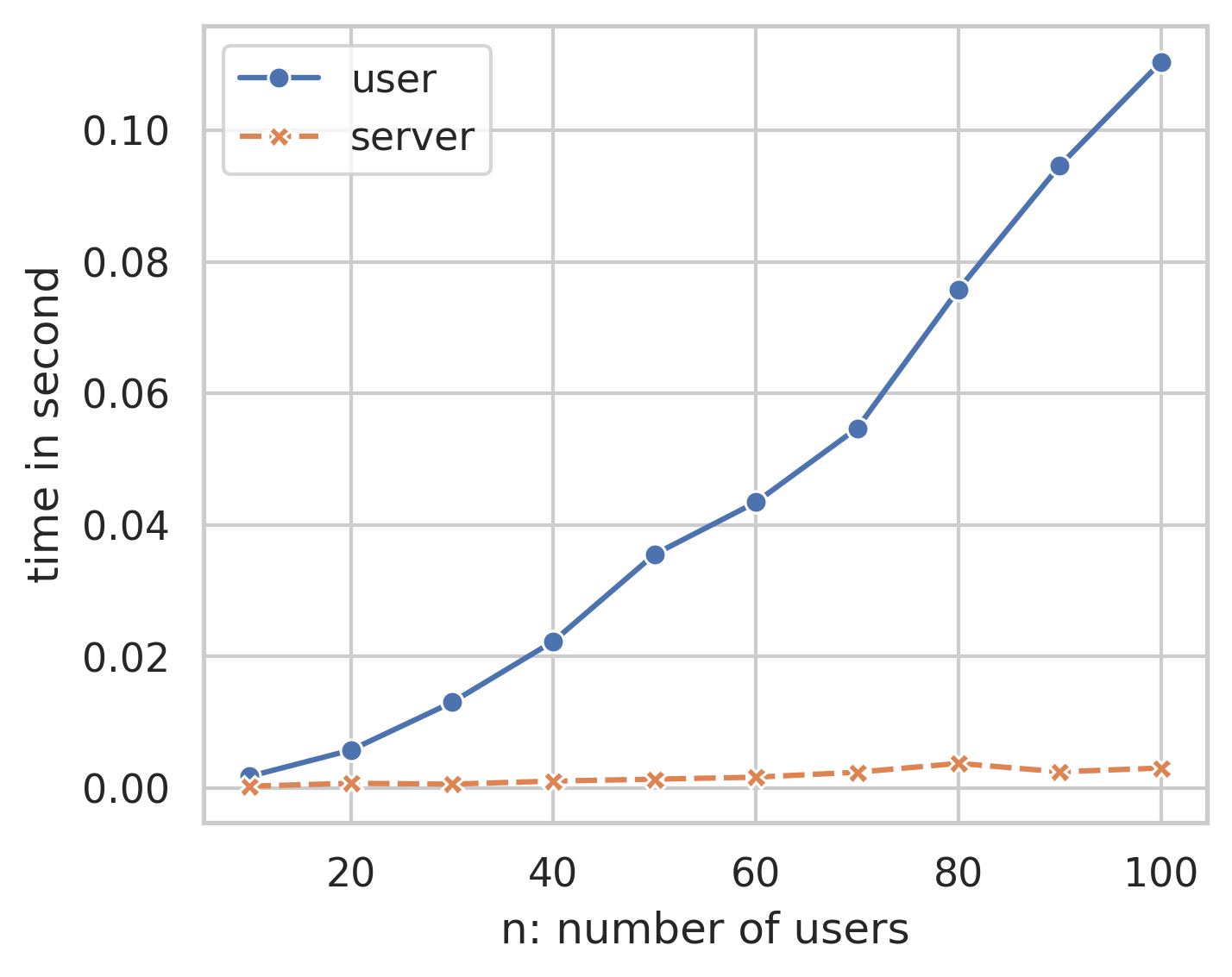}
        \caption{time vs. user number}
        \label{fig: time vs user num}
    \end{subfigure}
    \hfill
    \begin{subfigure}[b]{0.23\textwidth}
        \centering
        \includegraphics[width=\textwidth]{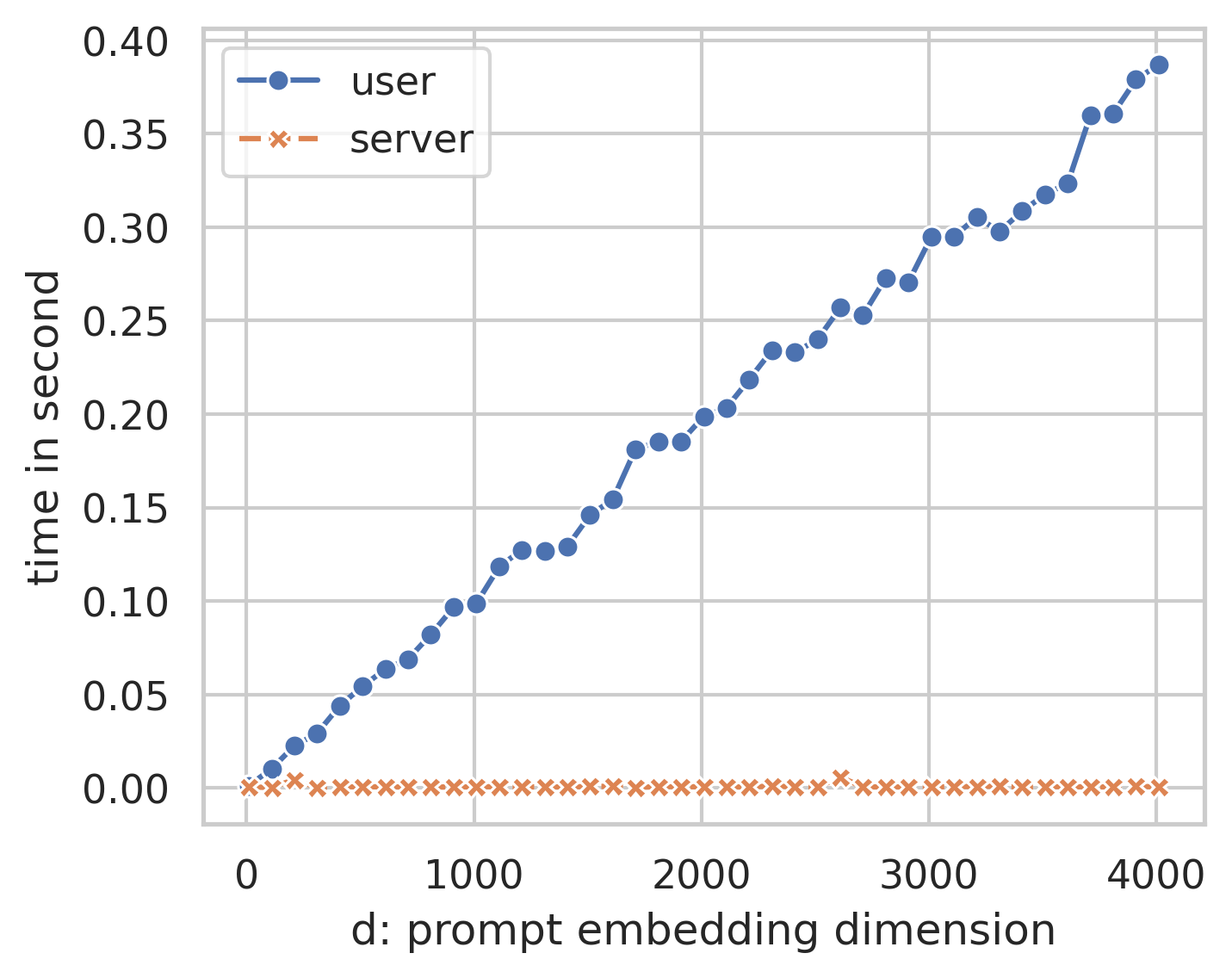}
        \caption{time vs. prompt dimension}
        \label{fig: time vs dim}
    \end{subfigure}
    \vspace{-3mm}
    \caption{The computational cost in seconds per round.}
    \label{fig: overhead}
    \vspace{-3mm}
\end{figure}

\begin{table}[tb]
  \caption{Communication/Computation Overheads of \scheme} 
  \vspace{-3mm}
  \label{tab: complexity}
  \begin{tabular}{ccc}
    \toprule
    \multicolumn{1}{c}{Communication} & \multicolumn{2}{c}{Computation} 
    \\\cmidrule(lr){1-1}\cmidrule(lr){2-3}
    User & User & Server \\
            \midrule
    $\mathcal{O}(\frac{nd}{\ell}+kn)$ & $\mathcal{O}(\frac{d}{\ell} n \log^{2}n + \frac{ndk+nd}{\ell})$ & $\mathcal{O}(kn^2\log^{2}n)$\\
  \bottomrule
\end{tabular}
\vspace{-3mm}
\end{table}

\noindent \textbf{User communication.} The communication overhead of each user consists of two parts: 1) the communication cost of secure secret sharing is $\mathcal{O}(\frac{nd}{\ell})$; 2) each client sends $\mathcal{O}(kn)$ distance shares to the server. The total communication cost of each client is $\mathcal{O}(\frac{nd}{\ell}+kn)$, which is dominated by $nd$. Here, $k$ is the number of cluster centers. 

\noindent \textbf{User computation.} The computation performed by each user consists of three parts: 1) utilizing fast polynomial interpolation and evaluation~\cite{kedlaya2011fast}, each user generates the secret shares of its local prompt with complexity 
$\mathcal{O}(\frac{d}{\ell} n \log^{2}n)$; 
2) Within each round, each user first updates the secret shares of $k$ centers, and computes the distances from each data point to each center in the finite domain, taking $\mathcal{O}(\frac{ndk}{\ell})$ operations. 3) Secure aggregation by $\mathcal{O}(\frac{nd}{\ell})$ The total computation complexity of each client
is $\mathcal{O}(\frac{d}{\ell} n \log^{2}n + \frac{ndk+nd}{\ell})$, which is dominated by $nd\log^{2}n$. 

\noindent \textbf{Server computation.} Server's computation complexity consists of two parts: 1) The decoding of the actual pair-wise distances takes $\mathcal{O}(kn^2\log^{2}n)$ operations; 2) The cost of running cluster assignment on each prompt is $\mathcal{O}(kn)$. Thus, the total computation complexity of the server over $s$ iterations is $\mathcal{O}(kn^2\log^{2}n)$.

Next, we present the empirical overheads of performing \secfc  in Figure \ref{fig: overhead} with dominant factors: $n$ and $d$. Privacy threshold $\alpha$ is set to $1/3$ consistently and $\ell = \left \lfloor\frac{n-t}{2}\right \rfloor$ accordingly. The large-enough prime $q$ is sampled in the scale of $10^{10}$ and the quantization parameter $\lambda$ is $10^3$ (see Appendix \ref{additional preli} for quantization details). As shown in Figure \ref{fig: overhead}, the computational cost is almost linearly predominated by $nd$ and the user's overhead is the primary cost. Though the server computation complexity has $n^2\log^2n$ term, its overhead is negligible in practice. The user's computation time reaches $0.4$s when the dimension is $4000$, however, when executing \scheme, \secfc uses reduced prompts with rank less than $10$, multiplying the number of tokens, the total dimension is only $d=150$. For communication costs, consider a $4$G network with $98$ Mbps bandwidth and the largest parameter settings in experiments. It takes $0.05$s for a user to share and less than $10^{-4}$s to communicate distances, which is even faster with networks like LAN or $5$G. Overall, the communication and computation overheads additionally for privacy preservation are negligible. 

\vspace{-2mm}
\section{Conclusion}

In this paper, we present a novel secure federated prompt personalization protocol, \scheme, that addresses multi-granular data heterogeneity via the hierarchical prompt adaption design and a privacy-preserving adaptive clustering mechanism, \secfc. Extensive experiments validate \scheme's robust performance across varying data unbalance, demonstrating consistent and superior performance over SOTA baselines. Moreover, \scheme theoretically guarantees and empirically validates strong privacy preservation of the user prompts, bridging the critical gap in the performance-privacy trade-off of existing FPP schemes.




\newpage
\appendix
\section*{Appendix}

\section{Additional Preliminaries}\label{additional preli}
In \scheme, the operations of data sharing and distance computation are carried out in a finite field $\mathbb{F}_q$, for some large prime $q$. Hence, for a data point ${\bf x}_i$ from the real field, one needs to first quantize it onto $\mathbb{F}_q$. We abuse the symbols only in this section.

\noindent {\bf Data Quantization.}\label{appendix:quant} The quantization technique in \scheme has been utilized in~\cite{so2020byzantine}~\cite{so2022lightsecagg}. To quantize a data point ${\bf x}_i$, we first scale it by $\lambda$, and embed the scaled value onto $\mathbb{F}_q$ such that

\begin{equation}
\label{quantization}
\bar{\mathbf{x}}_i = \mathcal{Q}(\mathbf{x}_i,\lambda)=\left\{\begin{array}{@{}ll}
\lfloor \lambda \mathbf{x}_i \rfloor ,   &\mathrm{if}~\mathbf{x}_i\geq 0  \\
\lfloor q+\lambda \mathbf{x}_i \rfloor, & \mathrm{if}~\mathbf{x}_i<0
\end{array}\right..
\end{equation}

Here $\lfloor x\rfloor$ denotes the largest integer less than or equal to $x$, and the quantization function $\mathcal{Q}$ is applied element-wise. Assuming each element of $\lambda {\bf x}_i$ is within $[-\eta, \eta)$ for some $\eta > 0$, then on the range of $\mathcal{Q}$, $\left [ 0,\eta \right )$ is the image of the positive part, and $\left [ q-\eta,q \right )$ is the image of the negative part. While $\lambda$ controls the precision loss of quantization, it also needs to be chosen so that overflow does not occur during computations of \scheme. A larger $\eta$ requires a larger finite field size $q$ to avoid computation overflow, and a smaller $\eta$ leads to a higher precision loss between $\mathbf{x}_i$ and $\bar{\mathbf{x}}_i$. We can choose a proper scaling factor $\lambda$ to preserve enough precision while keeping the field size practical. To avoid computation overflow, we should choose $q$ such that all the intermediate computation results on the scaled data $\lambda {\bf x}_i$ are within the range $\left ( -\frac{q}{2},\frac{q}{2} \right )$. In the worst case, the largest distance across data points and cluster centers  $D \triangleq \underset{i \in [m], h \in [k]}{\max} \left \| \boldsymbol{\mu}_{h}-\left |  \mathcal{S}_h \right | \cdot \mathbf{x}_{i} \right \|_2^2 $ results in the largest output value. Therefore, we should choose $q$ that is at least $2\lambda^2 D$. 

\noindent {\bf Lagrange Coded Computing (LCC).}\label{appendix:lcc} The technique used when each client secret shares its data with the other clients is Lagrange Coded Computing (LCC). As a result, each client gets a hold of the entire dataset in a coded manner, thereby being able to compute the secret shares of the distances between each pair of datapoints and cluster centers. Similar secret-sharing operations between clients have been widely utilized in FL literature to improve performance, privacy, and robustness (see e.g., \cite{shao2022dres}, \cite{schlegel2023codedpaddedfl}, \cite{so2020byzantine}, \cite{bonawitz2017practical}, \cite{so2020scalable}). In \cite{shao2022dres}, authors tackle the issue of non-iid data distribution of local client datasets, and use LCC-based secret sharing of client data to improve the convergence of FL models. In Reference \cite{schlegel2023codedpaddedfl}, authors use Shamir's secret sharing to encode and share their local datasets with the other clients. The added redundancy in the encoding process helps mitigate stragglers in FL, without sacrificing local data privacy. In Reference \cite{so2020byzantine}, clients exchange secret shares of their local model updates in a verifiable manner during FL iterations, so that each client can make sure that others have sent valid secret shares, which are then utilized by the server to decode the pairwise distances among client models for outlier detection and removal before aggregation. In \cite{bonawitz2017practical}, secret shares of the client models are utilized to handle client dropouts during secure model aggregation. Authors of \cite{so2020scalable} study the setting of decentralized FL, and utilize secret shares of private datasets among multiple parties to perform distributed training, while still preserving privacy.

\section{Supplementary to Proof}\label{appdix: supp proof}
$g_{m}(\cdot)$ and $h_n(\cdot)$ are two family of functions introduced in ~\cite{moser2020expected}. $g_{m}(\cdot): \mathbb{R}^+\rightarrow\mathbb{R}$ with $m\in \mathbb{N}$, which has the following expression:
\begin{equation}\scriptsize
g_m(\xi) \triangleq \begin{cases}\ln (\xi)-\mathrm{Ei}(-\xi)+\sum_{j=1}^{m-1}(-1)^{\prime}\left[e^{-\xi}(j-1)!-\frac{(m-1)!}{j(m-1-j)!}\right]\left(\frac{1}{\xi}\right)^{\prime} & \text { if } \xi>0, \\ \psi(m) & \text { if } \xi=0 .\end{cases}
\end{equation}
where $\mathrm{Ei}$ is the exponential integral and $\psi$ is digamma function. 
$h_n(\cdot): \mathbb{R}^+\rightarrow\mathbb{R}$ with $n\in \mathbb{N}^{odd}$, which has the following expression:
\begin{equation}\scriptsize
h_n(\xi) \triangleq \begin{cases}-\gamma-2 \ln (2)+2 \xi \cdot{ }_2 F_2\left(1,1 ; \frac{3}{2}, 2 ;-\xi\right) & \\ +\sum_{j=1}^{\frac{n-1}{2}}(-1)^{j-1} \Gamma\left(j-\frac{1}{2}\right) \cdot\left[\sqrt{\xi} e^{-\xi} \operatorname{erfi}(\sqrt{\xi})+\sum_{i=1}^{j-1} \frac{(-1)^i \xi^i}{\Gamma\left(i+\frac{1}{2}\right)}\right]\left(\frac{1}{\xi}\right)^j & \text { if } \xi>0, \\ \psi\left(\frac{n}{2}\right) & \text { if } \xi=0 .\end{cases}
\end{equation}
where ${ }_2 F_2$ is a generalized hypergeometric function and $\operatorname{erfi}$ is the imaginary error function.
For the analysis of the function families,  please refer to \cite{moser2020expected}.

\begin{proof} (Part II)
    In addition to Part I of the proof, we continue to discuss the other case when the prompt is outside of the given cluster. We continue from equation (\ref{eq: mi expansion to h}). When prompt $\overline{\mathbf{P}}_i$ is outside of the cluster, $h(D^2)$ is non-central $\chi^2$-distribution with $d$ degrees of freedom. The non-centrality parameter $\tau$ is defined by $\mu_i$. Hence, similar to equation (\ref{eq: tau}), $\tau_{i}=d\cdot\mu_i^2$ . Then:
    \begin{equation}\label{eq: h D outside}
        h(D^2) = \begin{cases}\ln (2)+h_d\left(\frac{\tau_i}{2}\right)  & \text { if } d \in \mathbb{N}^{\text {odd }}, \\\ln (2)+g_{d / 2}\left(\frac{\tau_i}{2}\right)  & \text { if } d \in \mathbb{N}^{\text {even }} .\end{cases}
    \end{equation}
    For the other term of $h\left(D^2\mid\overline{\mathbf{P}}_i\right)$, different from equation (\ref{eq: conditional entropy expantion}), we do not need to isolate $\overline{\mathbf{P}}_i$. Therefore, there is no constant term $c$ in the final expression of the conditional entropy. Combining equation (\ref{eq: h D outside}) and equation (\ref{eq: conditional entropy}) without $c$ yields the counterpart of the theorem. Note that if the prompt distribution has $\mu_i=0$, equation (\ref{eq: h D outside}) still reduces to the first part of the proof. 
\end{proof}



\begin{thebibliography}{71}
\bibliographystyle{ACM-Reference-Format}

\ifx \showCODEN    \undefined \def \showCODEN     #1{\unskip}     \fi
\ifx \showISBNx    \undefined \def \showISBNx     #1{\unskip}     \fi
\ifx \showISBNxiii \undefined \def \showISBNxiii  #1{\unskip}     \fi
\ifx \showISSN     \undefined \def \showISSN      #1{\unskip}     \fi
\ifx \showLCCN     \undefined \def \showLCCN      #1{\unskip}     \fi
\ifx \shownote     \undefined \def \shownote      #1{#1}          \fi
\ifx \showarticletitle \undefined \def \showarticletitle #1{#1}   \fi
\ifx \showURL      \undefined \def \showURL       {\relax}        \fi
\providecommand\bibfield[2]{#2}
\providecommand\bibinfo[2]{#2}
\providecommand\natexlab[1]{#1}
\providecommand\showeprint[2][]{arXiv:#2}

\bibitem[Abadi et~al\mbox{.}(2016)]%
        {abadi2016deep}
\bibfield{author}{\bibinfo{person}{Martin Abadi}, \bibinfo{person}{Andy Chu}, \bibinfo{person}{Ian Goodfellow}, \bibinfo{person}{H~Brendan McMahan}, \bibinfo{person}{Ilya Mironov}, \bibinfo{person}{Kunal Talwar}, {and} \bibinfo{person}{Li Zhang}.} \bibinfo{year}{2016}\natexlab{}.
\newblock \showarticletitle{Deep learning with differential privacy}. In \bibinfo{booktitle}{\emph{Proceedings of the 2016 ACM SIGSAC conference on computer and communications security}}. \bibinfo{pages}{308--318}.
\newblock


\bibitem[Bhatia et~al\mbox{.}(2004)]%
        {bhatia2004adaptive}
\bibfield{author}{\bibinfo{person}{Sanjiv~K Bhatia} {et~al\mbox{.}}} \bibinfo{year}{2004}\natexlab{}.
\newblock \showarticletitle{Adaptive K-Means Clustering.}. In \bibinfo{booktitle}{\emph{FLAIRS}}. \bibinfo{pages}{695--699}.
\newblock


\bibitem[Bonawitz et~al\mbox{.}(2017)]%
        {bonawitz2017practical}
\bibfield{author}{\bibinfo{person}{Keith Bonawitz}, \bibinfo{person}{Vladimir Ivanov}, \bibinfo{person}{Ben Kreuter}, \bibinfo{person}{Antonio Marcedone}, \bibinfo{person}{H~Brendan McMahan}, \bibinfo{person}{Sarvar Patel}, \bibinfo{person}{Daniel Ramage}, \bibinfo{person}{Aaron Segal}, {and} \bibinfo{person}{Karn Seth}.} \bibinfo{year}{2017}\natexlab{}.
\newblock \showarticletitle{Practical secure aggregation for privacy-preserving machine learning}. In \bibinfo{booktitle}{\emph{proceedings of the 2017 ACM SIGSAC Conference on Computer and Communications Security}}. \bibinfo{pages}{1175--1191}.
\newblock


\bibitem[Bossard et~al\mbox{.}(2014)]%
        {bossard14}
\bibfield{author}{\bibinfo{person}{Lukas Bossard}, \bibinfo{person}{Matthieu Guillaumin}, {and} \bibinfo{person}{Luc Van~Gool}.} \bibinfo{year}{2014}\natexlab{}.
\newblock \showarticletitle{Food-101 -- Mining Discriminative Components with Random Forests}. In \bibinfo{booktitle}{\emph{European Conference on Computer Vision}}.
\newblock


\bibitem[Buyukates et~al\mbox{.}(2024)]%
        {buyukates2024lightverifl}
\bibfield{author}{\bibinfo{person}{Baturalp Buyukates}, \bibinfo{person}{Jinhyun So}, \bibinfo{person}{Hessam Mahdavifar}, {and} \bibinfo{person}{Salman Avestimehr}.} \bibinfo{year}{2024}\natexlab{}.
\newblock \showarticletitle{LightVeriFL: A lightweight and verifiable secure aggregation for federated learning}.
\newblock \bibinfo{journal}{\emph{IEEE Journal on Selected Areas in Information Theory}} (\bibinfo{year}{2024}).
\newblock


\bibitem[Chen et~al\mbox{.}(2023)]%
        {chen2023unleashing}
\bibfield{author}{\bibinfo{person}{Banghao Chen}, \bibinfo{person}{Zhaofeng Zhang}, \bibinfo{person}{Nicolas Langren{\'e}}, {and} \bibinfo{person}{Shengxin Zhu}.} \bibinfo{year}{2023}\natexlab{}.
\newblock \showarticletitle{Unleashing the potential of prompt engineering in large language models: a comprehensive review}.
\newblock \bibinfo{journal}{\emph{arXiv preprint arXiv:2310.14735}} (\bibinfo{year}{2023}).
\newblock


\bibitem[Cimpoi et~al\mbox{.}(2014)]%
        {cimpoi14describing}
\bibfield{author}{\bibinfo{person}{M. Cimpoi}, \bibinfo{person}{S. Maji}, \bibinfo{person}{I. Kokkinos}, \bibinfo{person}{S. Mohamed}, \bibinfo{person}{}, {and} \bibinfo{person}{A. Vedaldi}.} \bibinfo{year}{2014}\natexlab{}.
\newblock \showarticletitle{Describing Textures in the Wild}. In \bibinfo{booktitle}{\emph{Proceedings of the {IEEE} Conf. on Computer Vision and Pattern Recognition ({CVPR})}}.
\newblock


\bibitem[Cui et~al\mbox{.}(2024)]%
        {cui2024harmonizing}
\bibfield{author}{\bibinfo{person}{Tianyu Cui}, \bibinfo{person}{Hongxia Li}, \bibinfo{person}{Jingya Wang}, {and} \bibinfo{person}{Ye Shi}.} \bibinfo{year}{2024}\natexlab{}.
\newblock \showarticletitle{Harmonizing Generalization and Personalization in Federated Prompt Learning}. In \bibinfo{booktitle}{\emph{International Conference on Machine Learning}}. PMLR, \bibinfo{pages}{9646--9661}.
\newblock


\bibitem[Darken and Moody(1990)]%
        {darken1990fast}
\bibfield{author}{\bibinfo{person}{Christian Darken} {and} \bibinfo{person}{John Moody}.} \bibinfo{year}{1990}\natexlab{}.
\newblock \showarticletitle{Fast adaptive k-means clustering: some empirical results}. In \bibinfo{booktitle}{\emph{1990 IJCNN international joint conference on neural networks}}. IEEE, \bibinfo{pages}{233--238}.
\newblock


\bibitem[Das et~al\mbox{.}(2025)]%
        {das2025security}
\bibfield{author}{\bibinfo{person}{Badhan~Chandra Das}, \bibinfo{person}{M~Hadi Amini}, {and} \bibinfo{person}{Yanzhao Wu}.} \bibinfo{year}{2025}\natexlab{}.
\newblock \showarticletitle{Security and privacy challenges of large language models: A survey}.
\newblock \bibinfo{journal}{\emph{Comput. Surveys}} \bibinfo{volume}{57}, \bibinfo{number}{6} (\bibinfo{year}{2025}), \bibinfo{pages}{1--39}.
\newblock


\bibitem[Demelius et~al\mbox{.}(2025)]%
        {demelius2025recent}
\bibfield{author}{\bibinfo{person}{Lea Demelius}, \bibinfo{person}{Roman Kern}, {and} \bibinfo{person}{Andreas Tr{\"u}gler}.} \bibinfo{year}{2025}\natexlab{}.
\newblock \showarticletitle{Recent advances of differential privacy in centralized deep learning: A systematic survey}.
\newblock \bibinfo{journal}{\emph{Comput. Surveys}} \bibinfo{volume}{57}, \bibinfo{number}{6} (\bibinfo{year}{2025}), \bibinfo{pages}{1--28}.
\newblock


\bibitem[Deng et~al\mbox{.}(2024)]%
        {deng2024unlocking}
\bibfield{author}{\bibinfo{person}{Wenlong Deng}, \bibinfo{person}{Christos Thrampoulidis}, {and} \bibinfo{person}{Xiaoxiao Li}.} \bibinfo{year}{2024}\natexlab{}.
\newblock \showarticletitle{Unlocking the potential of prompt-tuning in bridging generalized and personalized federated learning}. In \bibinfo{booktitle}{\emph{Proceedings of the IEEE/CVF Conference on Computer Vision and Pattern Recognition}}. \bibinfo{pages}{6087--6097}.
\newblock


\bibitem[Dosovitskiy et~al\mbox{.}(2020)]%
        {dosovitskiy2020image}
\bibfield{author}{\bibinfo{person}{Alexey Dosovitskiy}, \bibinfo{person}{Lucas Beyer}, \bibinfo{person}{Alexander Kolesnikov}, \bibinfo{person}{Dirk Weissenborn}, \bibinfo{person}{Xiaohua Zhai}, \bibinfo{person}{Thomas Unterthiner}, \bibinfo{person}{Mostafa Dehghani}, \bibinfo{person}{Matthias Minderer}, \bibinfo{person}{Georg Heigold}, \bibinfo{person}{Sylvain Gelly}, {et~al\mbox{.}}} \bibinfo{year}{2020}\natexlab{}.
\newblock \showarticletitle{An image is worth 16x16 words: Transformers for image recognition at scale}.
\newblock \bibinfo{journal}{\emph{arXiv preprint arXiv:2010.11929}} (\bibinfo{year}{2020}).
\newblock


\bibitem[Du et~al\mbox{.}(2024)]%
        {du2024sok}
\bibfield{author}{\bibinfo{person}{Jiacheng Du}, \bibinfo{person}{Jiahui Hu}, \bibinfo{person}{Zhibo Wang}, \bibinfo{person}{Peng Sun}, \bibinfo{person}{Neil~Zhenqiang Gong}, \bibinfo{person}{Kui Ren}, {and} \bibinfo{person}{Chun Chen}.} \bibinfo{year}{2024}\natexlab{}.
\newblock \showarticletitle{SoK: On Gradient Leakage in Federated Learning}.
\newblock \bibinfo{journal}{\emph{arXiv preprint arXiv:2404.05403}} (\bibinfo{year}{2024}).
\newblock


\bibitem[Du et~al\mbox{.}(2023)]%
        {du2023dp}
\bibfield{author}{\bibinfo{person}{Minxin Du}, \bibinfo{person}{Xiang Yue}, \bibinfo{person}{Sherman~SM Chow}, \bibinfo{person}{Tianhao Wang}, \bibinfo{person}{Chenyu Huang}, {and} \bibinfo{person}{Huan Sun}.} \bibinfo{year}{2023}\natexlab{}.
\newblock \showarticletitle{Dp-forward: Fine-tuning and inference on language models with differential privacy in forward pass}. In \bibinfo{booktitle}{\emph{Proceedings of the 2023 ACM SIGSAC Conference on Computer and Communications Security}}. \bibinfo{pages}{2665--2679}.
\newblock


\bibitem[Dwork(2006)]%
        {dwork2006differential}
\bibfield{author}{\bibinfo{person}{Cynthia Dwork}.} \bibinfo{year}{2006}\natexlab{}.
\newblock \showarticletitle{Differential privacy}. In \bibinfo{booktitle}{\emph{International colloquium on automata, languages, and programming}}. Springer, \bibinfo{pages}{1--12}.
\newblock


\bibitem[Edemacu and Wu(2024)]%
        {edemacu2024privacy}
\bibfield{author}{\bibinfo{person}{Kennedy Edemacu} {and} \bibinfo{person}{Xintao Wu}.} \bibinfo{year}{2024}\natexlab{}.
\newblock \showarticletitle{Privacy preserving prompt engineering: A survey}.
\newblock \bibinfo{journal}{\emph{Comput. Surveys}} (\bibinfo{year}{2024}).
\newblock


\bibitem[Feng et~al\mbox{.}(2024)]%
        {feng2024uncovering}
\bibfield{author}{\bibinfo{person}{Xinguo Feng}, \bibinfo{person}{Zhongkui Ma}, \bibinfo{person}{Zihan Wang}, \bibinfo{person}{Eu~Joe Chegne}, \bibinfo{person}{Mengyao Ma}, \bibinfo{person}{Alsharif Abuadbba}, {and} \bibinfo{person}{Guangdong Bai}.} \bibinfo{year}{2024}\natexlab{}.
\newblock \showarticletitle{Uncovering Gradient Inversion Risks in Practical Language Model Training}. In \bibinfo{booktitle}{\emph{Proceedings of the 2024 on ACM SIGSAC Conference on Computer and Communications Security}}. \bibinfo{pages}{3525--3539}.
\newblock


\bibitem[Gao et~al\mbox{.}(2018)]%
        {gao2018demystifying}
\bibfield{author}{\bibinfo{person}{Weihao Gao}, \bibinfo{person}{Sewoong Oh}, {and} \bibinfo{person}{Pramod Viswanath}.} \bibinfo{year}{2018}\natexlab{}.
\newblock \showarticletitle{Demystifying fixed $ k $-nearest neighbor information estimators}.
\newblock \bibinfo{journal}{\emph{IEEE Transactions on Information Theory}} \bibinfo{volume}{64}, \bibinfo{number}{8} (\bibinfo{year}{2018}), \bibinfo{pages}{5629--5661}.
\newblock


\bibitem[Geiping et~al\mbox{.}(2020)]%
        {geiping2020inverting}
\bibfield{author}{\bibinfo{person}{Jonas Geiping}, \bibinfo{person}{Hartmut Bauermeister}, \bibinfo{person}{Hannah Dr{\"o}ge}, {and} \bibinfo{person}{Michael Moeller}.} \bibinfo{year}{2020}\natexlab{}.
\newblock \showarticletitle{Inverting gradients-how easy is it to break privacy in federated learning?}
\newblock \bibinfo{journal}{\emph{Advances in neural information processing systems}}  \bibinfo{volume}{33} (\bibinfo{year}{2020}), \bibinfo{pages}{16937--16947}.
\newblock


\bibitem[Guo et~al\mbox{.}(2023a)]%
        {guo2023pfedprompt}
\bibfield{author}{\bibinfo{person}{Tao Guo}, \bibinfo{person}{Song Guo}, {and} \bibinfo{person}{Junxiao Wang}.} \bibinfo{year}{2023}\natexlab{a}.
\newblock \showarticletitle{Pfedprompt: Learning personalized prompt for vision-language models in federated learning}. In \bibinfo{booktitle}{\emph{Proceedings of the ACM Web Conference 2023}}. \bibinfo{pages}{1364--1374}.
\newblock


\bibitem[Guo et~al\mbox{.}(2023b)]%
        {guo2023promptfl}
\bibfield{author}{\bibinfo{person}{Tao Guo}, \bibinfo{person}{Song Guo}, \bibinfo{person}{Junxiao Wang}, \bibinfo{person}{Xueyang Tang}, {and} \bibinfo{person}{Wenchao Xu}.} \bibinfo{year}{2023}\natexlab{b}.
\newblock \showarticletitle{Promptfl: Let federated participants cooperatively learn prompts instead of models--federated learning in age of foundation model}.
\newblock \bibinfo{journal}{\emph{IEEE Transactions on Mobile Computing}} \bibinfo{volume}{23}, \bibinfo{number}{5} (\bibinfo{year}{2023}), \bibinfo{pages}{5179--5194}.
\newblock


\bibitem[Hou et~al\mbox{.}(2024)]%
        {hou2024priroagg}
\bibfield{author}{\bibinfo{person}{Sizai Hou}, \bibinfo{person}{Songze Li}, \bibinfo{person}{Tayyebeh Jahani-Nezhad}, {and} \bibinfo{person}{Giuseppe Caire}.} \bibinfo{year}{2024}\natexlab{}.
\newblock \showarticletitle{PriRoAgg: Achieving Robust Model Aggregation with Minimum Privacy Leakage for Federated Learning}.
\newblock \bibinfo{journal}{\emph{arXiv preprint arXiv:2407.08954}} (\bibinfo{year}{2024}).
\newblock


\bibitem[Jahani-Nezhad et~al\mbox{.}(2023)]%
        {jahani2023swiftagg}
\bibfield{author}{\bibinfo{person}{Tayyebeh Jahani-Nezhad}, \bibinfo{person}{Mohammad~Ali Maddah-Ali}, \bibinfo{person}{Songze Li}, {and} \bibinfo{person}{Giuseppe Caire}.} \bibinfo{year}{2023}\natexlab{}.
\newblock \showarticletitle{SwiftAgg+: Achieving asymptotically optimal communication loads in secure aggregation for federated learning}.
\newblock \bibinfo{journal}{\emph{IEEE Journal on Selected Areas in Communications}} \bibinfo{volume}{41}, \bibinfo{number}{4} (\bibinfo{year}{2023}), \bibinfo{pages}{977--989}.
\newblock


\bibitem[Jia et~al\mbox{.}(2022)]%
        {prompttuning1}
\bibfield{author}{\bibinfo{person}{Menglin Jia}, \bibinfo{person}{Luming Tang}, \bibinfo{person}{Bor-Chun Chen}, \bibinfo{person}{Claire Cardie}, \bibinfo{person}{Serge Belongie}, \bibinfo{person}{Bharath Hariharan}, {and} \bibinfo{person}{Ser-Nam Lim}.} \bibinfo{year}{2022}\natexlab{}.
\newblock \showarticletitle{Visual prompt tuning}. In \bibinfo{booktitle}{\emph{European conference on computer vision}}. Springer, \bibinfo{pages}{709--727}.
\newblock


\bibitem[Kedlaya and Umans(2011)]%
        {kedlaya2011fast}
\bibfield{author}{\bibinfo{person}{Kiran~S. Kedlaya} {and} \bibinfo{person}{Christopher Umans}.} \bibinfo{year}{2011}\natexlab{}.
\newblock \showarticletitle{Fast polynomial factorization and modular composition}.
\newblock \bibinfo{journal}{\emph{SIAM J. Comput.}} \bibinfo{volume}{40}, \bibinfo{number}{6} (\bibinfo{year}{2011}), \bibinfo{pages}{1767--1802}.
\newblock


\bibitem[Kraskov et~al\mbox{.}(2004)]%
        {kraskov2004KSG}
\bibfield{author}{\bibinfo{person}{Alexander Kraskov}, \bibinfo{person}{Harald St{\"o}gbauer}, {and} \bibinfo{person}{Peter Grassberger}.} \bibinfo{year}{2004}\natexlab{}.
\newblock \showarticletitle{Estimating mutual information}.
\newblock \bibinfo{journal}{\emph{Physical Review E—Statistical, Nonlinear, and Soft Matter Physics}} \bibinfo{volume}{69}, \bibinfo{number}{6} (\bibinfo{year}{2004}), \bibinfo{pages}{066138}.
\newblock


\bibitem[Krizhevsky et~al\mbox{.}(2009)]%
        {krizhevsky2009learning}
\bibfield{author}{\bibinfo{person}{Alex Krizhevsky}, \bibinfo{person}{Geoffrey Hinton}, {et~al\mbox{.}}} \bibinfo{year}{2009}\natexlab{}.
\newblock \showarticletitle{Learning multiple layers of features from tiny images}.
\newblock  (\bibinfo{year}{2009}).
\newblock


\bibitem[Lester et~al\mbox{.}(2021)]%
        {prompttuning2}
\bibfield{author}{\bibinfo{person}{Brian Lester}, \bibinfo{person}{Rami Al-Rfou}, {and} \bibinfo{person}{Noah Constant}.} \bibinfo{year}{2021}\natexlab{}.
\newblock \showarticletitle{The power of scale for parameter-efficient prompt tuning}.
\newblock \bibinfo{journal}{\emph{arXiv preprint arXiv:2104.08691}} (\bibinfo{year}{2021}).
\newblock


\bibitem[Li et~al\mbox{.}(2022a)]%
        {li_andreeto_ranzato_perona_2022}
\bibfield{author}{\bibinfo{person}{Fei-Fei Li}, \bibinfo{person}{Marco Andreeto}, \bibinfo{person}{Marc'Aurelio Ranzato}, {and} \bibinfo{person}{Pietro Perona}.} \bibinfo{year}{2022}\natexlab{a}.
\newblock \bibinfo{title}{Caltech 101}.
\newblock
\href{https://doi.org/10.22002/D1.20086}{doi:\nolinkurl{10.22002/D1.20086}}


\bibitem[Li et~al\mbox{.}(2023)]%
        {li2023visual}
\bibfield{author}{\bibinfo{person}{Guanghao Li}, \bibinfo{person}{Wansen Wu}, \bibinfo{person}{Yan Sun}, \bibinfo{person}{Li Shen}, \bibinfo{person}{Baoyuan Wu}, {and} \bibinfo{person}{Dacheng Tao}.} \bibinfo{year}{2023}\natexlab{}.
\newblock \showarticletitle{Visual prompt based personalized federated learning}.
\newblock \bibinfo{journal}{\emph{arXiv preprint arXiv:2303.08678}} (\bibinfo{year}{2023}).
\newblock


\bibitem[Li et~al\mbox{.}(2024)]%
        {li2024FedOTP}
\bibfield{author}{\bibinfo{person}{Hongxia Li}, \bibinfo{person}{Wei Huang}, \bibinfo{person}{Jingya Wang}, {and} \bibinfo{person}{Ye Shi}.} \bibinfo{year}{2024}\natexlab{}.
\newblock \showarticletitle{Global and local prompts cooperation via optimal transport for federated learning}. In \bibinfo{booktitle}{\emph{Proceedings of the IEEE/CVF Conference on Computer Vision and Pattern Recognition}}. \bibinfo{pages}{12151--12161}.
\newblock


\bibitem[Li et~al\mbox{.}(2022b)]%
        {li2022auditing}
\bibfield{author}{\bibinfo{person}{Zhuohang Li}, \bibinfo{person}{Jiaxin Zhang}, \bibinfo{person}{Luyang Liu}, {and} \bibinfo{person}{Jian Liu}.} \bibinfo{year}{2022}\natexlab{b}.
\newblock \showarticletitle{Auditing privacy defenses in federated learning via generative gradient leakage}. In \bibinfo{booktitle}{\emph{Proceedings of the IEEE/CVF Conference on Computer Vision and Pattern Recognition}}. \bibinfo{pages}{10132--10142}.
\newblock


\bibitem[Liu et~al\mbox{.}(2021)]%
        {prompttuning3}
\bibfield{author}{\bibinfo{person}{Xiao Liu}, \bibinfo{person}{Kaixuan Ji}, \bibinfo{person}{Yicheng Fu}, \bibinfo{person}{Weng~Lam Tam}, \bibinfo{person}{Zhengxiao Du}, \bibinfo{person}{Zhilin Yang}, {and} \bibinfo{person}{Jie Tang}.} \bibinfo{year}{2021}\natexlab{}.
\newblock \showarticletitle{P-tuning v2: Prompt tuning can be comparable to fine-tuning universally across scales and tasks}.
\newblock \bibinfo{journal}{\emph{arXiv preprint arXiv:2110.07602}} (\bibinfo{year}{2021}).
\newblock


\bibitem[Lu et~al\mbox{.}(2022)]%
        {prompttuning5}
\bibfield{author}{\bibinfo{person}{Yuning Lu}, \bibinfo{person}{Jianzhuang Liu}, \bibinfo{person}{Yonggang Zhang}, \bibinfo{person}{Yajing Liu}, {and} \bibinfo{person}{Xinmei Tian}.} \bibinfo{year}{2022}\natexlab{}.
\newblock \showarticletitle{Prompt distribution learning}. In \bibinfo{booktitle}{\emph{Proceedings of the IEEE/CVF Conference on Computer Vision and Pattern Recognition}}. \bibinfo{pages}{5206--5215}.
\newblock


\bibitem[MacQueen(1967)]%
        {macqueen1967some}
\bibfield{author}{\bibinfo{person}{James MacQueen}.} \bibinfo{year}{1967}\natexlab{}.
\newblock \showarticletitle{Some methods for classification and analysis of multivariate observations}. In \bibinfo{booktitle}{\emph{Proceedings of the Fifth Berkeley Symposium on Mathematical Statistics and Probability, Volume 1: Statistics}}, Vol.~\bibinfo{volume}{5}. University of California press, \bibinfo{pages}{281--298}.
\newblock


\bibitem[Moser(2020)]%
        {moser2020expected}
\bibfield{author}{\bibinfo{person}{Stefan~M Moser}.} \bibinfo{year}{2020}\natexlab{}.
\newblock \showarticletitle{Expected logarithm and negative integer moments of a noncentral $\chi$ 2-distributed random variable}.
\newblock \bibinfo{journal}{\emph{Entropy}} \bibinfo{volume}{22}, \bibinfo{number}{9} (\bibinfo{year}{2020}), \bibinfo{pages}{1048}.
\newblock


\bibitem[Nilsback and Zisserman(2008)]%
        {nilsback2008automated}
\bibfield{author}{\bibinfo{person}{Maria-Elena Nilsback} {and} \bibinfo{person}{Andrew Zisserman}.} \bibinfo{year}{2008}\natexlab{}.
\newblock \showarticletitle{Automated flower classification over a large number of classes}. In \bibinfo{booktitle}{\emph{2008 Sixth Indian conference on computer vision, graphics \& image processing}}. IEEE, \bibinfo{pages}{722--729}.
\newblock


\bibitem[OpenAI({[n.\,d.]})]%
        {gptstore}
\bibfield{author}{\bibinfo{person}{OpenAI}.} \bibinfo{year}{[n.\,d.]}\natexlab{}.
\newblock \bibinfo{booktitle}{\emph{OpenAI GPT Store}}.
\newblock
\urldef\tempurl%
\url{https://openai.com/index/introducing-the-gpt-store/}
\showURL{%
\tempurl}


\bibitem[Parkhi et~al\mbox{.}(2012)]%
        {parkhi2012cats}
\bibfield{author}{\bibinfo{person}{Omkar~M Parkhi}, \bibinfo{person}{Andrea Vedaldi}, \bibinfo{person}{Andrew Zisserman}, {and} \bibinfo{person}{CV Jawahar}.} \bibinfo{year}{2012}\natexlab{}.
\newblock \showarticletitle{Cats and dogs}. In \bibinfo{booktitle}{\emph{2012 IEEE conference on computer vision and pattern recognition}}. IEEE, \bibinfo{pages}{3498--3505}.
\newblock


\bibitem[Petrov et~al\mbox{.}(2024)]%
        {petrov2024dager}
\bibfield{author}{\bibinfo{person}{Ivo Petrov}, \bibinfo{person}{Dimitar~I Dimitrov}, \bibinfo{person}{Maximilian Baader}, \bibinfo{person}{Mark M{\"u}ller}, {and} \bibinfo{person}{Martin Vechev}.} \bibinfo{year}{2024}\natexlab{}.
\newblock \showarticletitle{Dager: Exact gradient inversion for large language models}.
\newblock \bibinfo{journal}{\emph{Advances in Neural Information Processing Systems}}  \bibinfo{volume}{37} (\bibinfo{year}{2024}), \bibinfo{pages}{87801--87830}.
\newblock


\bibitem[PromptBase({[n.\,d.]})]%
        {PromptBase}
\bibfield{author}{\bibinfo{person}{PromptBase}.} \bibinfo{year}{[n.\,d.]}\natexlab{}.
\newblock \bibinfo{booktitle}{\emph{PromptBase}}.
\newblock
\urldef\tempurl%
\url{https://promptbase.com}
\showURL{%
\tempurl}


\bibitem[Radford et~al\mbox{.}(2021)]%
        {radford2021learning}
\bibfield{author}{\bibinfo{person}{Alec Radford}, \bibinfo{person}{Jong~Wook Kim}, \bibinfo{person}{Chris Hallacy}, \bibinfo{person}{Aditya Ramesh}, \bibinfo{person}{Gabriel Goh}, \bibinfo{person}{Sandhini Agarwal}, \bibinfo{person}{Girish Sastry}, \bibinfo{person}{Amanda Askell}, \bibinfo{person}{Pamela Mishkin}, \bibinfo{person}{Jack Clark}, {et~al\mbox{.}}} \bibinfo{year}{2021}\natexlab{}.
\newblock \showarticletitle{Learning transferable visual models from natural language supervision}. In \bibinfo{booktitle}{\emph{International conference on machine learning}}. PmLR, \bibinfo{pages}{8748--8763}.
\newblock


\bibitem[Ross(2014)]%
        {ross2014mutual}
\bibfield{author}{\bibinfo{person}{Brian~C Ross}.} \bibinfo{year}{2014}\natexlab{}.
\newblock \showarticletitle{Mutual information between discrete and continuous data sets}.
\newblock \bibinfo{journal}{\emph{PloS one}} \bibinfo{volume}{9}, \bibinfo{number}{2} (\bibinfo{year}{2014}), \bibinfo{pages}{e87357}.
\newblock


\bibitem[Schlegel et~al\mbox{.}(2023)]%
        {schlegel2023codedpaddedfl}
\bibfield{author}{\bibinfo{person}{Reent Schlegel}, \bibinfo{person}{Siddhartha Kumar}, \bibinfo{person}{Eirik Rosnes}, {and} \bibinfo{person}{Alexandre~Graell i Amat}.} \bibinfo{year}{2023}\natexlab{}.
\newblock \showarticletitle{CodedPaddedFL and CodedSecAgg: Straggler mitigation and secure aggregation in federated learning}.
\newblock \bibinfo{journal}{\emph{IEEE Transactions on Communications}} (\bibinfo{year}{2023}).
\newblock


\bibitem[Shamir(1979)]%
        {shamir1979share}
\bibfield{author}{\bibinfo{person}{Adi Shamir}.} \bibinfo{year}{1979}\natexlab{}.
\newblock \showarticletitle{How to share a secret}.
\newblock \bibinfo{journal}{\emph{Commun. ACM}} \bibinfo{volume}{22}, \bibinfo{number}{11} (\bibinfo{year}{1979}), \bibinfo{pages}{612--613}.
\newblock


\bibitem[Shao et~al\mbox{.}(2022)]%
        {shao2022dres}
\bibfield{author}{\bibinfo{person}{Jiawei Shao}, \bibinfo{person}{Yuchang Sun}, \bibinfo{person}{Songze Li}, {and} \bibinfo{person}{Jun Zhang}.} \bibinfo{year}{2022}\natexlab{}.
\newblock \showarticletitle{Dres-fl: Dropout-resilient secure federated learning for non-iid clients via secret data sharing}.
\newblock \bibinfo{journal}{\emph{Advances in Neural Information Processing Systems}}  \bibinfo{volume}{35} (\bibinfo{year}{2022}), \bibinfo{pages}{10533--10545}.
\newblock


\bibitem[Shen et~al\mbox{.}(2024)]%
        {shen2024prompt}
\bibfield{author}{\bibinfo{person}{Xinyue Shen}, \bibinfo{person}{Yiting Qu}, \bibinfo{person}{Michael Backes}, {and} \bibinfo{person}{Yang Zhang}.} \bibinfo{year}{2024}\natexlab{}.
\newblock \showarticletitle{Prompt Stealing Attacks Against $\{$Text-to-Image$\}$ Generation Models}. In \bibinfo{booktitle}{\emph{33rd USENIX Security Symposium (USENIX Security 24)}}. \bibinfo{pages}{5823--5840}.
\newblock


\bibitem[Shi et~al\mbox{.}(2022)]%
        {shi2022just}
\bibfield{author}{\bibinfo{person}{Weiyan Shi}, \bibinfo{person}{Ryan Shea}, \bibinfo{person}{Si Chen}, \bibinfo{person}{Chiyuan Zhang}, \bibinfo{person}{Ruoxi Jia}, {and} \bibinfo{person}{Zhou Yu}.} \bibinfo{year}{2022}\natexlab{}.
\newblock \showarticletitle{Just fine-tune twice: Selective differential privacy for large language models}.
\newblock \bibinfo{journal}{\emph{arXiv preprint arXiv:2204.07667}} (\bibinfo{year}{2022}).
\newblock


\bibitem[Shu et~al\mbox{.}(2022)]%
        {prompttuning4}
\bibfield{author}{\bibinfo{person}{Manli Shu}, \bibinfo{person}{Weili Nie}, \bibinfo{person}{De-An Huang}, \bibinfo{person}{Zhiding Yu}, \bibinfo{person}{Tom Goldstein}, \bibinfo{person}{Anima Anandkumar}, {and} \bibinfo{person}{Chaowei Xiao}.} \bibinfo{year}{2022}\natexlab{}.
\newblock \showarticletitle{Test-time prompt tuning for zero-shot generalization in vision-language models}.
\newblock \bibinfo{journal}{\emph{Advances in Neural Information Processing Systems}}  \bibinfo{volume}{35} (\bibinfo{year}{2022}), \bibinfo{pages}{14274--14289}.
\newblock


\bibitem[snackprompt({[n.\,d.]})]%
        {snackprompt}
\bibfield{author}{\bibinfo{person}{snackprompt}.} \bibinfo{year}{[n.\,d.]}\natexlab{}.
\newblock \bibinfo{booktitle}{\emph{snackprompt}}.
\newblock
\urldef\tempurl%
\url{https://snackprompt.com}
\showURL{%
\tempurl}


\bibitem[So et~al\mbox{.}(2020a)]%
        {so2020byzantine}
\bibfield{author}{\bibinfo{person}{Jinhyun So}, \bibinfo{person}{Ba{\c{s}}ak G{\"u}ler}, {and} \bibinfo{person}{A~Salman Avestimehr}.} \bibinfo{year}{2020}\natexlab{a}.
\newblock \showarticletitle{Byzantine-resilient secure federated learning}.
\newblock \bibinfo{journal}{\emph{IEEE Journal on Selected Areas in Communications}} \bibinfo{volume}{39}, \bibinfo{number}{7} (\bibinfo{year}{2020}), \bibinfo{pages}{2168--2181}.
\newblock


\bibitem[So et~al\mbox{.}(2020b)]%
        {so2020scalable}
\bibfield{author}{\bibinfo{person}{Jinhyun So}, \bibinfo{person}{Basak Guler}, {and} \bibinfo{person}{Salman Avestimehr}.} \bibinfo{year}{2020}\natexlab{b}.
\newblock \showarticletitle{A scalable approach for privacy-preserving collaborative machine learning}.
\newblock \bibinfo{journal}{\emph{Advances in Neural Information Processing Systems}}  \bibinfo{volume}{33} (\bibinfo{year}{2020}), \bibinfo{pages}{8054--8066}.
\newblock


\bibitem[So et~al\mbox{.}(2022)]%
        {so2022lightsecagg}
\bibfield{author}{\bibinfo{person}{Jinhyun So}, \bibinfo{person}{Chaoyang He}, \bibinfo{person}{Chien-Sheng Yang}, \bibinfo{person}{Songze Li}, \bibinfo{person}{Qian Yu}, \bibinfo{person}{Ramy E~Ali}, \bibinfo{person}{Basak Guler}, {and} \bibinfo{person}{Salman Avestimehr}.} \bibinfo{year}{2022}\natexlab{}.
\newblock \showarticletitle{Lightsecagg: a lightweight and versatile design for secure aggregation in federated learning}.
\newblock \bibinfo{journal}{\emph{Proceedings of Machine Learning and Systems}}  \bibinfo{volume}{4} (\bibinfo{year}{2022}), \bibinfo{pages}{694--720}.
\newblock


\bibitem[Tran et~al\mbox{.}(2025)]%
        {tran2025privacy}
\bibfield{author}{\bibinfo{person}{Linh Tran}, \bibinfo{person}{Wei Sun}, \bibinfo{person}{Stacy Patterson}, {and} \bibinfo{person}{Ana Milanova}.} \bibinfo{year}{2025}\natexlab{}.
\newblock \showarticletitle{Privacy-Preserving Personalized Federated Prompt Learning for Multimodal Large Language Models}.
\newblock \bibinfo{journal}{\emph{arXiv preprint arXiv:2501.13904}} (\bibinfo{year}{2025}).
\newblock


\bibitem[Vu et~al\mbox{.}(2024)]%
        {vu2024analysis}
\bibfield{author}{\bibinfo{person}{Minh Vu}, \bibinfo{person}{Truc Nguyen}, \bibinfo{person}{My~T Thai}, {et~al\mbox{.}}} \bibinfo{year}{2024}\natexlab{}.
\newblock \showarticletitle{Analysis of privacy leakage in federated large language models}. In \bibinfo{booktitle}{\emph{International Conference on Artificial Intelligence and Statistics}}. PMLR, \bibinfo{pages}{1423--1431}.
\newblock


\bibitem[Wang et~al\mbox{.}(2021)]%
        {wang2021privacy}
\bibfield{author}{\bibinfo{person}{Binghui Wang}, \bibinfo{person}{Jiayi Guo}, \bibinfo{person}{Ang Li}, \bibinfo{person}{Yiran Chen}, {and} \bibinfo{person}{Hai Li}.} \bibinfo{year}{2021}\natexlab{}.
\newblock \showarticletitle{Privacy-preserving representation learning on graphs: A mutual information perspective}. In \bibinfo{booktitle}{\emph{Proceedings of the 27th acm sigkdd conference on knowledge discovery \& data mining}}. \bibinfo{pages}{1667--1676}.
\newblock


\bibitem[Wei et~al\mbox{.}(2020)]%
        {wei2020federated}
\bibfield{author}{\bibinfo{person}{Kang Wei}, \bibinfo{person}{Jun Li}, \bibinfo{person}{Ming Ding}, \bibinfo{person}{Chuan Ma}, \bibinfo{person}{Howard~H Yang}, \bibinfo{person}{Farhad Farokhi}, \bibinfo{person}{Shi Jin}, \bibinfo{person}{Tony~QS Quek}, {and} \bibinfo{person}{H~Vincent Poor}.} \bibinfo{year}{2020}\natexlab{}.
\newblock \showarticletitle{Federated learning with differential privacy: Algorithms and performance analysis}.
\newblock \bibinfo{journal}{\emph{IEEE transactions on information forensics and security}}  \bibinfo{volume}{15} (\bibinfo{year}{2020}), \bibinfo{pages}{3454--3469}.
\newblock


\bibitem[Wold et~al\mbox{.}(1987)]%
        {wold1987principal}
\bibfield{author}{\bibinfo{person}{Svante Wold}, \bibinfo{person}{Kim Esbensen}, {and} \bibinfo{person}{Paul Geladi}.} \bibinfo{year}{1987}\natexlab{}.
\newblock \showarticletitle{Principal component analysis}.
\newblock \bibinfo{journal}{\emph{Chemometrics and intelligent laboratory systems}} \bibinfo{volume}{2}, \bibinfo{number}{1-3} (\bibinfo{year}{1987}), \bibinfo{pages}{37--52}.
\newblock


\bibitem[Wu et~al\mbox{.}(2024)]%
        {wu2024quantifying}
\bibfield{author}{\bibinfo{person}{Yixin Wu}, \bibinfo{person}{Rui Wen}, \bibinfo{person}{Michael Backes}, \bibinfo{person}{Pascal Berrang}, \bibinfo{person}{Mathias Humbert}, \bibinfo{person}{Yun Shen}, {and} \bibinfo{person}{Yang Zhang}.} \bibinfo{year}{2024}\natexlab{}.
\newblock \showarticletitle{Quantifying privacy risks of prompts in visual prompt learning}. In \bibinfo{booktitle}{\emph{33rd USENIX Security Symposium (USENIX Security 24)}}. \bibinfo{pages}{5841--5858}.
\newblock


\bibitem[Xia et~al\mbox{.}(2020)]%
        {xia2020fast}
\bibfield{author}{\bibinfo{person}{Shuyin Xia}, \bibinfo{person}{Daowan Peng}, \bibinfo{person}{Deyu Meng}, \bibinfo{person}{Changqing Zhang}, \bibinfo{person}{Guoyin Wang}, \bibinfo{person}{Elisabeth Giem}, \bibinfo{person}{Wei Wei}, {and} \bibinfo{person}{Zizhong Chen}.} \bibinfo{year}{2020}\natexlab{}.
\newblock \showarticletitle{A fast adaptive k-means with no bounds}.
\newblock \bibinfo{journal}{\emph{IEEE Transactions on Pattern Analysis and Machine Intelligence}} (\bibinfo{year}{2020}).
\newblock


\bibitem[Xiao et~al\mbox{.}(2025)]%
        {xiao2025differential}
\bibfield{author}{\bibinfo{person}{Xingpeng Xiao}, \bibinfo{person}{Yaomin Zhang}, \bibinfo{person}{Heyao Chen}, \bibinfo{person}{Wenkun Ren}, \bibinfo{person}{Junyi Zhang}, {and} \bibinfo{person}{Jian Xu}.} \bibinfo{year}{2025}\natexlab{}.
\newblock \showarticletitle{A Differential Privacy-Based Mechanism for Preventing Data Leakage in Large Language Model Training}.
\newblock \bibinfo{journal}{\emph{Academic Journal of Sociology and Management}} \bibinfo{volume}{3}, \bibinfo{number}{2} (\bibinfo{year}{2025}), \bibinfo{pages}{33--42}.
\newblock


\bibitem[Yang et~al\mbox{.}(2023b)]%
        {yang2023efficient}
\bibfield{author}{\bibinfo{person}{Fu-En Yang}, \bibinfo{person}{Chien-Yi Wang}, {and} \bibinfo{person}{Yu-Chiang~Frank Wang}.} \bibinfo{year}{2023}\natexlab{b}.
\newblock \showarticletitle{Efficient model personalization in federated learning via client-specific prompt generation}. In \bibinfo{booktitle}{\emph{Proceedings of the IEEE/CVF International Conference on Computer Vision}}. \bibinfo{pages}{19159--19168}.
\newblock


\bibitem[Yang et~al\mbox{.}(2023a)]%
        {yang2023gradient}
\bibfield{author}{\bibinfo{person}{Haomiao Yang}, \bibinfo{person}{Mengyu Ge}, \bibinfo{person}{Dongyun Xue}, \bibinfo{person}{Kunlan Xiang}, \bibinfo{person}{Hongwei Li}, {and} \bibinfo{person}{Rongxing Lu}.} \bibinfo{year}{2023}\natexlab{a}.
\newblock \showarticletitle{Gradient leakage attacks in federated learning: Research frontiers, taxonomy and future directions}.
\newblock \bibinfo{journal}{\emph{IEEE Network}} (\bibinfo{year}{2023}).
\newblock


\bibitem[Yu et~al\mbox{.}(2019)]%
        {yu2019lagrange}
\bibfield{author}{\bibinfo{person}{Qian Yu}, \bibinfo{person}{Songze Li}, \bibinfo{person}{Netanel Raviv}, \bibinfo{person}{Seyed Mohammadreza~Mousavi Kalan}, \bibinfo{person}{Mahdi Soltanolkotabi}, {and} \bibinfo{person}{Salman~A. Avestimehr}.} \bibinfo{year}{2019}\natexlab{}.
\newblock \showarticletitle{Lagrange coded computing: Optimal design for resiliency, security, and privacy}. In \bibinfo{booktitle}{\emph{The 22nd International Conference on Artificial Intelligence and Statistics}}. PMLR, \bibinfo{pages}{1215--1225}.
\newblock


\bibitem[Zhang et~al\mbox{.}(2024)]%
        {zhang2024graphleak}
\bibfield{author}{\bibinfo{person}{Xi~Sheryl Zhang}, \bibinfo{person}{Weifan Guan}, \bibinfo{person}{Jiahao Lu}, \bibinfo{person}{Zhaopeng Qiu}, \bibinfo{person}{Jian Cheng}, \bibinfo{person}{Xian Wu}, {and} \bibinfo{person}{Yefeng Zheng}.} \bibinfo{year}{2024}\natexlab{}.
\newblock \showarticletitle{GraphLeak: Patient Record Leakage through Gradients with Knowledge Graph}. In \bibinfo{booktitle}{\emph{Proceedings of the ACM Web Conference 2024}}. \bibinfo{pages}{4706--4716}.
\newblock


\bibitem[Zhao et~al\mbox{.}(2020)]%
        {zhao2020idlg}
\bibfield{author}{\bibinfo{person}{Bo Zhao}, \bibinfo{person}{Konda~Reddy Mopuri}, {and} \bibinfo{person}{Hakan Bilen}.} \bibinfo{year}{2020}\natexlab{}.
\newblock \showarticletitle{idlg: Improved deep leakage from gradients}.
\newblock \bibinfo{journal}{\emph{arXiv preprint arXiv:2001.02610}} (\bibinfo{year}{2020}).
\newblock


\bibitem[Zhao et~al\mbox{.}(2023)]%
        {zhao2023fedprompt}
\bibfield{author}{\bibinfo{person}{Haodong Zhao}, \bibinfo{person}{Wei Du}, \bibinfo{person}{Fangqi Li}, \bibinfo{person}{Peixuan Li}, {and} \bibinfo{person}{Gongshen Liu}.} \bibinfo{year}{2023}\natexlab{}.
\newblock \showarticletitle{Fedprompt: Communication-efficient and privacy-preserving prompt tuning in federated learning}. In \bibinfo{booktitle}{\emph{ICASSP 2023-2023 IEEE International Conference on Acoustics, Speech and Signal Processing (ICASSP)}}. IEEE, \bibinfo{pages}{1--5}.
\newblock


\bibitem[Zhou et~al\mbox{.}(2022a)]%
        {zhou2022conditional}
\bibfield{author}{\bibinfo{person}{Kaiyang Zhou}, \bibinfo{person}{Jingkang Yang}, \bibinfo{person}{Chen~Change Loy}, {and} \bibinfo{person}{Ziwei Liu}.} \bibinfo{year}{2022}\natexlab{a}.
\newblock \showarticletitle{Conditional prompt learning for vision-language models}. In \bibinfo{booktitle}{\emph{Proceedings of the IEEE/CVF conference on computer vision and pattern recognition}}. \bibinfo{pages}{16816--16825}.
\newblock


\bibitem[Zhou et~al\mbox{.}(2022b)]%
        {zhou2022learning}
\bibfield{author}{\bibinfo{person}{Kaiyang Zhou}, \bibinfo{person}{Jingkang Yang}, \bibinfo{person}{Chen~Change Loy}, {and} \bibinfo{person}{Ziwei Liu}.} \bibinfo{year}{2022}\natexlab{b}.
\newblock \showarticletitle{Learning to prompt for vision-language models}.
\newblock \bibinfo{journal}{\emph{International Journal of Computer Vision}} \bibinfo{volume}{130}, \bibinfo{number}{9} (\bibinfo{year}{2022}), \bibinfo{pages}{2337--2348}.
\newblock


\bibitem[Zhu et~al\mbox{.}(2019)]%
        {zhu2019deep}
\bibfield{author}{\bibinfo{person}{Ligeng Zhu}, \bibinfo{person}{Zhijian Liu}, {and} \bibinfo{person}{Song Han}.} \bibinfo{year}{2019}\natexlab{}.
\newblock \showarticletitle{Deep leakage from gradients}.
\newblock \bibinfo{journal}{\emph{Advances in neural information processing systems}}  \bibinfo{volume}{32} (\bibinfo{year}{2019}).
\newblock


\end{thebibliography}


\end{document}